# Topological near fields generated by topological structures

Jie Peng[1,*], Ruo-Yang Zhang[2,*], Shiqi Jia[1], Wei Liu[3], and Shubo Wang[1,4,†]

[1]*Department of Physics, City University of Hong Kong, Tat Chee Avenue, Kowloon, Hong Kong, China*
[2]*Department of Physics, The Hong Kong University of Science and Technology, Clear Water Bay, Kowloon, Hong Kong, China*
[3]*College for Advanced Interdisciplinary Studies, National University of Defense Technology, Changsha, Hunan 410073, China*
[4]*City University of Hong Kong Shenzhen Research Institute, Shenzhen, Guangdong 518057, China*

* These authors contributed equally.
† Correspondence should be addressed to: Shubo Wang (shubwang@cityu.edu.hk)

## Abstract

The central idea of metamaterials and metaoptics is that, besides their base materials, the geometry of structures offers a broad extra dimension to explore for exotic functionalities. Here, we discover that the topology of structures fundamentally dictates the topological properties of optical near fields and offers a new dimension to exploit for optical functionalities that are irrelevant to specific material constituents or structural geometries. We find that the nontrivial topology of metal structures ensures the birth of polarization singularities (PSs) in the near field with rich morphologies and intriguing spatial evolutions including merging, bifurcation, and topological transition. By mapping the PSs to non-Hermitian exceptional points and employing homotopy theory, we extract the core invariant that governs the topological classification of the PSs and the conservation law that regulates their spatial evolutions. The results have effectively bridged three vibrant fields of singular optics, topological photonics, and non-Hermitian physics, with potential applications in chiral sensing, chiral quantum optics, and beyond photonics in other wave systems.

The concept of topology has provided new perspectives for physicists to explore unconventional properties of physical systems, such as the one-way edge states in topological insulators and their counterparts in classical wave systems that have attracted considerable interest recently[1,2]. These properties are associated with the topology of the momentum space (i.e., Brillouin zone). Topology in the real space can also give rise to intriguing physical properties and phenomena. In particular, Maxwell's equations allow unique solutions with nontrivial topology in the real space, where the field lines, phase singularities (i.e., optical vortices), or polarization disclinations can form links and knots[3–12]. These topological configurations of electromagnetic fields can enable highly flexible manipulations of phase and polarization with unprecedented precision for various applications.

At an arbitrary point of the three-dimensional (3D) real space, the end of electric/magnetic field vector of a generic monochromatic electromagnetic wave traces out an ellipse, i.e., the field is elliptically polarized[13]. The distribution of the polarization ellipses can form topological defects known as polarization singularities (PSs)[14], which include C points (where the field is circularly polarized and the direction of the major axis of polarization ellipse is ill-defined), L points (where the field is linearly polarized and the normal direction of polarization ellipse is ill-defined), as well as V points (where the field norm is zero and the field direction is ill-defined). The 3D lines formed of the PSs are referred to as C lines, L lines, and V lines accordingly. These polarization singularity lines (PSLs) can emerge during light focusing[9,15], scattering[16–20], interference[10,21], in nanostructures including metasurfaces[22] and photonic crystals[23]. The integration of singularity and topology theory has revealed richer physics including geometric phases[24,25], bound states in the continuum (BIC)[26–28], Hermitian topological nodal degeneracies[29,30], and non-Hermitian exceptional points (EPs)[31–35], etc.

The explicit geometry of optical structures decides the local resonance of optical modes and gives rise to novel optical devices and wave-functional materials, such as nanoantennas[36], metamaterials[37], and metasurfaces[38]. As a result, conventional studies in photonics mainly focus on the geometry and rarely pay attention to the overall topology of the structures investigated. It thus becomes interesting to ask: What optical properties are determined solely by the overall topology of optical structures? How can the topology of optical polarization fields and the topology of optical structures be interconnected?

In this article, we establish a universal and exact connection between the topology of optical fields and that of optical structures, revealing how the births and topological evolutions of magnetic PSs in the near fields are bounded by the topology and symmetry of the structures. As is required by the Poincaré-Hopf theorem, the existence of the PSs is topologically protected, and they are characterized by quantized topological indices with the index sum solely decided by the genuses of the structures. We further demonstrate that, by incorporating extra spatial symmetries (such as the mirror symmetry and the generalized rotational symmetry) for the structures and the incident fields, higher-order PSs/PSLs and topologically stable nexuses of PSLs can emerge, around which the polarizations evolve into striking configurations such as mirror-symmetric double-twist Möbius strips[39–41]. To grasp the underlying invariant properties of the continuous evolutions of PSs (e.g., merging, bifurcation, and topological transition), we map the real-space C points to the EPs of a 2-band non-Hermitian Hamiltonian, upon which homotopy theory can be directly employed to identify the core invariant to classify all topological evolutions.

## Results

**Polarization singularities protected by surface topology of structures**

A general monochromatic magnetic field can be expressed as $\mathbf{H} = (\mathbf{A} + i\mathbf{B})e^{i\theta}$, where $\mathbf{A}$ and $\mathbf{B}$ are the major and minor axes of the polarization ellipse, and $\theta =$ $\frac{\arg(\mathbf{H}\cdot\mathbf{H})}{2}$ is a proper phase. The C points of the magnetic field $\mathbf{H}$ correspond to the phase singularities of the scalar field $\Psi = \mathbf{H}\cdot\mathbf{H}$ and generally can form stable C lines in 3D space without any symmetry protection. A C line can be characterized by two topological indices (i.e. winding numbers): the polarization index $I_{\mathrm{pl}} = \frac{1}{4\pi}\oint d\phi$ and the phase index $I_{\mathrm{ph}} = \frac{1}{2\pi}\oint \nabla\varphi \cdot d\mathbf{r}$, where $\phi$ is the azimuthal angle on the Poincaré sphere for the in-plane polarization (i.e., in the plane of the polarization ellipses at a C point) and $\varphi = \mathrm{Arg}(\Psi)$, and both integrals are evaluated on a small loop enclosing the C line[25,42]. We note that $I_{\mathrm{pl}}$ is uniquely defined at each C point provided that the magnetic spin $\mathbf{S} = \mathrm{Im}(\mathbf{H}^* \times \mathbf{H})$ is not normal to the C line. At the points where $\mathbf{S}$ is normal to the C line, $I_{\mathrm{pl}}$ may change sign, making $I_{\mathrm{pl}}$ not a global invariant along the C line. In contrast, although the sign of $I_{\mathrm{ph}}$ depends on the direction of the integration loop, it is invariant against continuously moving the loop along the C line. Therefore, $I_{\mathrm{ph}}$ can endow the C line with a positive direction $\mathbf{t}_c = \mathrm{sign}(I_{\mathrm{ph}})\mathbf{t}$, where $\mathbf{t}$ is the tangent vector of the C line complying with the right-hand rule of the integration loop of $I_{\mathrm{ph}}$ [43]. One can prove that the two indices are related by $I_{\mathrm{pl}} = \mathrm{sign}(\mathbf{t}\cdot\mathbf{S})I_{\mathrm{ph}}/2$ (see Supplementary Note 2).

Let us first consider a metal sphere under the incidence of a plane wave propagating in $z$ direction and the magnetic field is linearly polarized in $y$ direction. Without loss of generality, we assume that the metal is gold characterized by a Drude model (see Methods). We conducted full-wave numerical simulations of the system by using a finite-element package COMSOL Multiphysics. The numerically obtained PSLs at the frequency $f = 100$ THz are shown in Fig. 1(a). Interestingly, a pair of C lines emerge near the surface of the sphere. For each C line, the polarization index $I_{\mathrm{pl}}$ changes from $+1/2$ (red) to $-1/2$ (blue) and then back to $+1/2$. Such sign flips happen at the points where $\mathbf{S} \perp \mathbf{t}_c$ (see Supplementary Note 2). In Fig.

1(e) and 1(f), we show the polarization ellipses on a plane perpendicular to **S** for $I_{pl} = +\frac{1}{2}$ and $I_{pl} = -\frac{1}{2}$, respectively, where the C points are marked by the red and blue dots, and the line segments inside the polarization ellipses denote the major axis **A**. The phase $\text{Arg}(\Psi)$ is also shown on the same plane, which clearly show a phase topological index $I_{ph} = +1$ and $I_{ph} = -1$, correspondingly, following the relationship $I_{pl} = \frac{I_{ph}}{2}$. The emergence of such a configuration of C lines is puzzling, and it becomes even intriguing for two coupled metal spheres, as shown in Fig. 1(b). Under the incidence of the same plane wave, two C lines with $I_{pl} = +\frac{1}{2}$ appear near the surfaces and connect the two spheres. In addition, a V line (magenta-colored) appears in the gap region between the spheres due to the accidental degeneracy of a pair of C lines with the same polarization index $I_{pl} = +\frac{1}{2}$. We thus can define its index as $I_{pl} = +\frac{1}{2} \times 2 = +1$ by summing up $I_{pl}$ of the two superposed C lines (see Supplementary Note 2). Although a V line cannot stably exist in general, the intersections of two C lines can form stable V points protected by mirror symmetry, as will be proved later.

We now ask the question: what determines the morphologies of the PSLs and their topological indices? It turns out that the answer is rooted in the topological properties associated with the geometry of the metal spheres. Under the excitation of the incident electromagnetic field, currents are induced in metal structures. These currents mainly localize within a thin surface layer of the structures, the thickness of which is approximately equal to the "skin depth" (i.e., the depth that light penetrates the metal)[44]. By the boundary conditions, the magnetic field **H** near the metal surface is dominated by its tangent component (see Supplementary Note 1). Consequently, the major axis **A** of the polarization ellipses can be considered a tangent bi-vector field (i.e., line field) defined on a two-dimensional (2D) smooth manifold $M$ (i.e., the surfaces of the structures). According to the Poincaré-Hopf theorem for tangent line fields, a finite number of isolated singularities of this field must emerge on the surfaces of the structures[45]. These singularities are exactly the C points and/or V points, and the sum of their topological polarization indices must satisfy $\sum_{\mathbf{r}_s} I_{\mathrm{pl}}(\mathbf{r}_s) = \chi(M)$, where $\mathbf{r}_s$ denote the PSs on the smooth manifold $M$, and $\chi$ denotes the Euler characteristic of $M$. For a closed orientable surface, $\chi$ is directly given by the genus $g$ of the surface: $\chi = 2 - 2g$[45]. For the single sphere case in Fig. 1(a), there are four C points on the sphere surface with the same index $I_{\mathrm{pl}} = +\frac{1}{2}$, thus, $\sum I_{\mathrm{pl}} = +\frac{1}{2} \times 4 = +2 = \chi$, consistent with the Poincaré-Hopf theorem. For the case of coupled spheres in Fig. 1(b), there are four C points and two V points on the spheres' surface with total index $\sum I_{\mathrm{pl}} = +\frac{1}{2} \times 4 + 1 \times 2 = +4 = \chi$, again satisfying the theorem.

To further verify the above interpretations based on Poincaré-Hopf theorem, we calculated the PSLs generated by metal structures with genus $g = 1$ and $g = 2$ as shown in Fig. 1(c) and (d), respectively. The same incident plane wave is applied in Fig. 1(a)-(d), and we focus on the frequency range with small skin depth. Under this condition, the physics is general and is not restricted to a particular frequency. For the single torus in Fig. 1(c), in addition to two C lines, we observe an accidental V line with $I_{\mathrm{pl}} = -1$ (green-colored) emerged inside the hole, corresponding to a pair of degenerate C lines with $I_{\mathrm{pl}} = -\frac{1}{2}$. The index sum of the PSLs for the single torus is $\sum I_{\mathrm{pl}} = +\frac{1}{2} \times 4 + (-1) \times 2 = 0$, which is equal to the Euler characteristic of a torus. Figure 1(d) shows the PSLs generated by the double-torus with genus $g = 2$. Remarkably, a rich structure of the PSLs appear with crossings of C lines. There are totally twenty C points on the metal surface with total indices $\sum I_{\mathrm{pl}} = +\frac{1}{2} \times 8 + \left(-\frac{1}{2}\right) \times 12 = -2$, again agreeing with the Euler characteristic of the double-torus. We have also simulated the case of triple-torus ($g = 3$) and the results also satisfy the theorem (data not shown). These results, for the first time, establish a direct relation between the topology of optical structures and the topological properties of optical near fields. This relation holds for arbitrary metal structures as long as their geometric surfaces are smooth and the skin depth is small.

Since the emergence of PSLs is protected by the topology of the structures, their global topological properties are robust against continuously varying the geometry of structures unless singular perturbations are introduced to the geometry. For example, if we introduce sharp edges into the sphere such that the radii of curvature near the edges are comparable with the skin depth, the total index of PSLs can change, as shown in Fig. 2(a), where $\sum I_{\text{pl}} = 0$. The presence of sharp edges (where the tangent planes and thus the tangent fields are not well defined) renders the original Poincaré-Hopf theorem for smooth manifolds inapplicable to the surface. Thus, the total index is not necessarily determined by the genus of the geometry. On the other hand, if the sharp edge is "smoothed out", as shown in Fig. 2(b), the global topological properties of the PSLs recover, i.e., the indices satisfy the theorem again. A second example is given in Fig. 2(c), where a cylindrical portion is removed from the sphere. In this case, the sharp edge induces two additional C lines with $I_{\text{pl}} = -\frac{1}{2}$, and the total index of the C lines is zero. After smoothing the edges, the total index recovers the value of $+2$, as shown in Fig. 2(d).

**Mapping to non-Hermitian exceptional points**

The PSLs evolve as they extend away from the surfaces of the structures, leading to merging, bifurcation, and topological transition in the 3D space. To understand these phenomena, we employ a mapping from the C points (real-space singularities) to the non-Hermitian EPs (parameter-space singularities). We introduce an auxiliary 2-band non-Hermitian Hamiltonian associated with the magnetic field, $\mathcal{H}(\mathbf{r}) = \mathbf{H}(\mathbf{r}) \cdot \vec{\sigma}$, where $\vec{\sigma} = (\hat{\sigma}_x, \hat{\sigma}_y, \hat{\sigma}_z)$ is the vector of Pauli matrices. The Hamiltonian has two eigenvalues $h_\pm = \pm\sqrt{\mathbf{H}\cdot\mathbf{H}}$ that are proportional to the eccentricity of the polarization ellipse $\left(1 - \frac{B^2}{A^2}\right)$ and are degenerate when the discriminant of the Hamiltonian's characteristic polynomial reduces to zero: $D = (h_+ - h_-)^2 = 4\mathbf{H}\cdot\mathbf{H} = 4\Psi = 0$, i.e., the condition for the emergence of a C point or a V point. Thus, the degeneracies of the non-Hermitian Hamiltonian, i.e., EPs and non-defective nodal points, just correspond to the C and V points of the magnetic field, respectively. This remarkable property allows a mapping from the topology of EPs given by the 2-band non-Hermitian Hamiltonian to the topology of C points in magnetic field. By borrowing the topological classification of the 2-band non-Hermitian Hamiltonian with separable bands[46], we can obtain the topological classification of the PSLs in the 3D real space. Specifically, in the absence of any symmetry constraint, the configuration space of the non-Hermitian Hamiltonian without degeneracies can be expressed by the coset space[46] $X \simeq (S^2 \times S^1)/\mathbb{Z}_2$ (see Methods), which can be physically understood as the configuration space of the magnetic field in the real space: $S^2$ stands for the orientation sphere of the major axis $\mathbf{A}$ of the polarization ellipse; $S^1$ corresponds to the circle of the phase $\varphi = \arg(\Psi)$; $\mathbb{Z}_2$ denotes the redundancy that both the direction of $\mathbf{A}$ and $e^{\frac{i\varphi}{2}}$ change sign simultaneously. We note that since only C and V points are of our interest, the vanishing of the minor axis $\mathbf{B}$ of the polarization ellipse (i.e., the condition of L points) is irrelevant to the topological classification here. The first homotopy group of the configuration space $X$ gives the topological classification of the polarization field along any closed loop in the space[46]: $\pi_1(X) = \pi_1\left(\frac{S^2 \times S^1}{\mathbb{Z}_2}\right) = \pi_1\left(\frac{S^1}{\mathbb{Z}_2}\right) = \mathbb{Z}$, which essentially classifies the topologically different morphisms of the PSLs (i.e., C lines in general) encircled by the loop. The $\mathbb{Z}$ topological invariant distinguishing different homotopy equivalence classes in $\pi_1(X)$ is just the phase winding number of $\Psi$ along a loop $\gamma$, i.e., phase index $I_{\text{ph}}(\gamma)$. It corresponds to the energy vorticity, also known as the discriminant number, in the notation of non-Hermitian physics[33–35]. Therefore, the phase index $I_{ph}(\gamma)$ is not only conserved against continuous deformation of the loop, but also a complete index that can characterize all topological phases associated with C lines. In comparison, the polarization index defined along an arbitrary loop, characterized by the trivial or Möbius twists of the major axis **A**, is not a complete index (see Supplementary Note 2). As we have shown previously, if a loop only encloses one C line, the phase index $I_{ph}$ defined with the loop can assign a positive direction $\mathbf{t}_c$ to the C line. More broadly, the topological index $I_{ph}(\gamma)$ defined with an arbitrary loop $\gamma$ counts the net number of directed C lines passing through the loop.

Since the topological invariant $\mathbb{Z}$ is equal to the phase index $I_{ph}$, we can apply this invariant index to understand the topological transition of PSLs. Consider the double-sphere case in Fig. 1(b), at large separation of the spheres, the PSLs must reduce to that of two isolated spheres (corresponding to two copies of Fig. 1(a)). This involves a topological transition. Figure 3(a)-(d) shows the PSLs for the double-sphere case when the separation is increased, where we label the C lines by 1, 2, 3, 4. Figure 3(e)-(h) show the phase $\varphi = \text{Arg}(\Psi)$ on the *yoz*-plane in the middle of the two spheres for the cases of Fig. 3(a)-(d). To approach the configuration in Fig. 3(d), the V line with $I_{ph} = 0$ in Fig. 3(a) must bifurcate into two C lines with $I_{ph} = \pm 1$ that can merge with the rest C lines. This is indeed the configuration in Fig. 3(b) and 3(f), where the V line bifurcates into two C lines 3 and 4 with phase indices $I_{ph} = +1$ and $I_{ph} = -1$, respectively. Merging and annihilation can only happen to a pair of C lines with opposite phase indices. This explains why C line 1 merges with C line 3, while C line 2 merges with C line 4, enabling the topological transition and opening a gap, as shown in Fig. 3(c), (d), (g) and (h). Additionally, the results in Fig. 3 indicate that C lines with opposite polarization index $I_{pl}$ do not necessarily annihilate because their chirality (i.e., spin **S**) can be different. An example is the C lines 1 and 4 in Fig. 3(b) and (f), which have opposite $I_{\text{pl}}$ and opposite chirality. On the other hand, this can be easily understood based on the phase index since both C lines carry $I_{\text{ph}} = -1$.

**Integration of topology and mirror symmetry**

To further understand the properties of the PSLs in Fig. 1, it is necessary to discuss the combined effect of mirror symmetry and topology. If we impose a *y*-mirror symmetry about the mirror plane $y = 0$, marked as $\Pi$, it is straightforward to show that the 2-band Hamiltonian satisfies $\mathcal{H}(\mathbf{r}) = \left(-\widehat{m}_y \mathbf{H}(\widehat{m}_y \mathbf{r})\right) \cdot \vec{\sigma} = \hat{\sigma}_y \mathcal{H}(\widehat{m}_y \mathbf{r}) \hat{\sigma}_y$, where $\widehat{m}_y = \text{diag}(1, -1, 1)$ denotes the mirror reflection operator for polar vectors about $y = 0$. Thus, the magnetic field in a mirror-symmetric system can be mapped to a Hamiltonian with a mirror symmetry. In this case, the magnetic field on the mirror plane $\Pi$ only has perpendicular component, $\mathbf{H}(x, 0, z) = H_y(x, 0, z)\hat{\mathbf{y}}$, hence the configuration space $X_\Pi$ of the nonsingular magnetic field on $\Pi$ is given by $X_\Pi = \{\mathbf{H} = H_y \hat{\mathbf{y}} \mid H_y \neq 0\} \simeq \mathbb{C} - 0 \simeq S^1$. Therefore, we obtain the topological classification of the magnetic fields along the loops in $\Pi$ according to the first homotopy group: $\pi_1(X_\Pi) = \pi_1(S^1) = \mathbb{Z}$, which is characterized by the different phase winding number of $H_y$ along the loops. In this case, topologically stable V points just appear at the phase singularities of $H_y$ in the mirror plane. The phase index of each V point defined on a loop encircling the V point in the mirror plane must be $I_{\text{ph}} = \frac{1}{2\pi} \oint \nabla \text{Arg}(H_y^2) \cdot d\mathbf{r} = \frac{1}{\pi} \oint \nabla \text{Arg}(H_y) \cdot d\mathbf{r} = \pm 2$. The conservation of the phase index $I_{\text{ph}} = \pm 2$ indicates that such a mirror-symmetry-protected V point is not an isolated singularity but manifests as the intersection point of two C lines that pierce the mirror plane from the same side (see Fig. 4).

We now apply the above results to understand the V points labeled as V1 and V2 in Fig. 1(d), which are nexuses of two mirror-partner C lines protected by a $y$-mirror symmetry of the system. Figure 4(a) shows the polarization ellipses and the phase $\mathrm{Arg}(\Psi)$ near V1 on a cutting plane and the $y$-mirror plane, respectively. The polarization ellipses indicate that the two C lines have opposite chirality $\mathbf{S}(\hat{m}_y \mathbf{r}) = -\hat{m}_y \mathbf{S}(\mathbf{r})$ (denoted by the blue and red colors of the ellipses) but same polarization index $I_{\mathrm{pl}} = +1/2$ (denoted by the red color of the C lines), as guaranteed by the mirror symmetry (see Supplementary Note 3). As a result, the two C lines are oppositely oriented about the mirror plane, i.e., $\mathbf{t}_c(\mathbf{r}_c) = -\mathbf{t}_c(\hat{m}_y \mathbf{r}_c)$ according to the relation $\mathbf{t}_c = \mathrm{sign}(I_{\mathrm{pl}}\mathbf{S} \cdot \mathbf{t})\mathbf{t}$, as shown by the yellow arrows on the C lines. In other words, both $\mathbf{S}$ and $\mathbf{t}_c$ behave as pseudo-vectors under mirror reflection. The phase $\mathrm{Arg}(\Psi)$ on the $y$-mirror plane has a $-4\pi$ variation around the V1 point, corresponding to a phase index $I_{\mathrm{ph}} = -2$ (viewed from $+\hat{\mathbf{y}}$ direction) and in accordance with the fact that the two C lines both point inward ($-\hat{\mathbf{y}}$ direction) from the mirror plane at the V1 point. Figure 4(b) shows the polarization ellipses and the phase $\mathrm{Arg}(\Psi)$ near the point V2 on a cutting plane and the $y$-mirror plane, respectively. Similar to the case of Fig. 4(a), the two C lines have opposite chirality but the same polarization index $I_{\mathrm{pl}} = -1/2$ and both puncture inwardly through the mirror plane at the V2 point, which is consistent with the phase index $I_{\mathrm{ph}} = -2$ around the V2 point as shown by $\mathrm{Arg}(\Psi)$ on the mirror plane.

It is known that interesting Möbius strips of polarizations appear around a single C line[8,9,39–41,47–49]. Generally, a polarization strip with odd number of twists must be topologically nontrivial, while a stripe with even number of twists can always be deformed to a cylinder with trivially aligned polarizations [25]. An interesting question is: Could mirror symmetry enrich the topological classification and give rise to nontrivial polarization structures around the nexus of C lines? To explore this question, let us consider the polarization major axes $\mathbf{A}$ on a transverse self-mirror symmetric (TSMS) loop (i.e., a loop that intersects with and is symmetric about the *y*-mirror plane). Since the two mirror-partner C lines have opposite directions, the phase index for this loop vanishes, which seems to indicate a trivial topology on the loop with boring spatial structure of **A**. However, under the constraint of mirror symmetry, the topological classification along such TSMS loops is determined by the relative homotopy group $\pi_1(X, X_\Pi) = \pi_1\left(\frac{S^2 \times S^1}{\mathbb{Z}_2}, S^1\right) = \mathbb{Z}_2$ (see Supplementary Note 3). Consequently, despite carrying zero phase index, the TSMS loops can still have a nontrivial $\mathbb{Z}_2$ topology manifesting as a mirror-symmetric double-twist Möbius strip of the polarization major axes **A**. Concretely, the major axis **A** on the mirror plane must point in the $\pm y$ direction due to the *y*-mirror symmetry. The evolution of **A** on the TSMS loops gives rise to only two possible cases: trivial and nontrivial topological loops. Along a trivial loop, the polarizations can be continuously deformed into an untwisted strip. In contrast, the polarizations along a nontrivial loop must be twisted once at each side of the mirror plane and hence form a mirror-symmetric double-twist Möbius strip protected by mirror symmetry. We note that the absolute times of winding of 3D polarizations are not a topological invariant, which can change arbitrary even number of times by deforming either the loop or the system without breaking the symmetry[25,39]. Thus, here the number of twists should be counted in a topologically stable sense (i.e., the minimal number of twists that cannot be untied by continuous deformation). In addition, a mirror-symmetry-protected double-twist (trivial) polarization strip must enclose odd (even) number of C lines at each side of the mirror plane (see Supplementary Note 3). The above analysis is confirmed by the numerical results in Fig. 4, where the pink arrows denote the polarization major axes **A** on a TSMS loop for the V1 and V2 cases. We notice the mirror-symmetric double-twist Möbius strips in both cases. This also explains why the two oppositely directed C lines at the two sides of the mirror plane cannot annihilate but must form a V point when meeting on the mirror plane.

Once the mirror symmetry is broken, the double-twist Möbius strip can be deformed into a trivial strip with no twist, and the two C lines will be gapped at the V point. Therefore, our study reveals that spatial symmetry can engender intriguing topologically nontrivial polarization configurations that cannot stably exist in the non-symmetric case.

**Integration of topology and generalized rotational symmetry**

Higher-order PSs/PSLs are generally unstable without symmetry protection and can easily transform into multiple lowest-order PSs/PSLs under perturbations[50]. The V points in Fig. 4 are a type of higher-order PSs protected by mirror symmetry. It is interesting to explore whether higher-order C points/lines can appear if the topology of structures is integrated with rotational symmetry.

We consider the metal sphere under the excitation of a right-handed circularly polarized plane wave $\mathbf{H}_{\text{in}} = H_0(\hat{\mathbf{z}} + i\hat{\mathbf{x}})e^{iky}$ which transforms under rotation about the $y$ axis as $R(\phi)\mathbf{H}_{\text{in}}(R(-\phi)\mathbf{r}) = e^{-i\phi}\mathbf{H}_{\text{in}}(\mathbf{r})$[51,52], where $R(\phi)$ denotes the rotation matrix with rotation angle $\phi$. The rotation matrix $R(\phi)$ has three eigenvalues $e^{-i\phi}$, $e^{i\phi}$ and 1 with the corresponding eigenvectors $\mathbf{R} = \frac{1}{\sqrt{2}}(\hat{\mathbf{z}} + i\hat{\mathbf{x}})$, $\mathbf{L} = \frac{1}{\sqrt{2}}(\hat{\mathbf{z}} - i\hat{\mathbf{x}})$, and $\hat{\mathbf{y}}$. Since the structure is axially symmetric about the $y$ axis, the total magnetic field should be invariant against continuous rotation about the $y$ axis up to a certain additional phase $R(\phi)\mathbf{H}(R(-\phi)\mathbf{r}) = e^{-i\phi}\mathbf{H}(\mathbf{r})$, which we dubbed as the generalized cylindrical symmetry $\bar{C}_\infty$ in order to differentiate it from the ordinary rotational symmetry (i.e., the system returns to itself without the additional phase after rotation). Along the $y$ axis, the magnetic field satisfies $R(\phi)\mathbf{H}(y\hat{\mathbf{y}}) = e^{-i\phi}\mathbf{H}(y\hat{\mathbf{y}})$, implying that $\mathbf{H}$ is the circularly polarized eigenvector of $R(\phi)$ with the same chirality as the incident field, i.e., $\mathbf{H} \propto \mathbf{R} = \frac{1}{\sqrt{2}}(\hat{\mathbf{z}} + i\hat{\mathbf{x}})$, and hence the magnetic spin $\mathbf{S}$ is fixed in the $+\hat{\mathbf{y}}$ direction. Moreover, since the scalar field satisfies $\Psi(R(\phi)\mathbf{r}) = $ $e^{2i\phi}\Psi(\mathbf{r})$, the phase index around the $y$ axis is quantized as $I_{\text{ph}} = \frac{1}{2\pi}\oint d\phi\, \partial_\phi \text{Arg}[\Psi(R(\phi)\mathbf{r})] = +2$, and accordingly $I_{\text{pl}} = \text{sign}(\mathbf{S}\cdot\hat{\mathbf{y}})I_{\text{ph}}/2 = +1$, indicating the total field along the $y$ axis forms a second-order C line. This prediction is confirmed by the numerical result in Fig. 5(a), where the red line threading the sphere corresponds to a C line with index $I_{\text{pl}} = +1$, and the index sum on the surface of the sphere remains unchanged, i.e., $\sum I_{\text{pl}} = +1 \times 2 = +2$. Similar phenomenon also happens in the single torus case, as shown in Fig. 5(b). However, the vanished Euler characteristic ($\chi = 0$) of the torus makes the C line with $I_{\text{pl}} = +1$ thread the hole, without touching the metal surface. It is worth noting that when the incident field is left-handed polarized $\mathbf{H}_{\text{in}} \propto \mathbf{L}$, $I_{\text{ph}}$ and $\mathbf{S}$ will reverse their sign simultaneously, while the polarization index of the C line will be left intact, i.e., $I_{\text{pl}} = +1$, as the result of the only possible configuration of a cylindrically symmetric line field vortex (see Fig. S4 in Supplementary Note 4).

When the structure illuminated by the circularly polarized wave $\mathbf{H}_{\text{in}}$ has a discrete rotational symmetry $C_n$ about the $y$ axis, the total magnetic field respects the generalized rotational symmetry $\bar{C}_n$: $R\left(\frac{2\pi}{n}\right)\mathbf{H}\left(R\left(-\frac{2\pi}{n}\right)\mathbf{r}\right) = e^{-\frac{i2\pi}{n}}\mathbf{H}(\mathbf{r})$. The discrete $\bar{C}_n$ symmetry can also protect that the total field forms a C line with $\mathbf{H} \propto \mathbf{R}$ along the $y$ aixs when $n \geq 3$ (for $n = 1, 2$, since the two eigenvalues $e^{\frac{i2\pi}{n}}, e^{-\frac{i2\pi}{n}}$ of $R\left(\frac{2\pi}{n}\right)$ are degenerate, $\mathbf{H}$ can be arbitrarily polarized along the $y$ aixs, and there is no C line on the $y$ axis), and the order of the C line depends on the order $n$ of the rotational symmetry. Concretely, since the $\bar{C}_n$ symmetric scalar field satisfies $\Psi\left(R\left(\frac{2\pi}{n}\right)\mathbf{r}\right) = e^{\frac{i4\pi}{n}}\Psi(\mathbf{r})$, the phase index along a circle, $c$, symmetric about the $y$ axis is $n\mathbb{Z}$ quantized: $I_{\text{ph}}(c) = 2 + nm$ (see Supplementary Note 4 for a rigorous proof), where $m$ is an arbitrary integer. The first term of $I_{\text{ph}}(c)$ (i.e., constant

2) corresponds to the phase index of systems respecting the generalized cylindrical symmetry $\bar{C}_\infty$ and hence must also obey all discrete $\bar{C}_n$ symmetries. The second term is due to the fact that the $\bar{C}_n$ symmetry ensures that the C lines must enter or exit the circle in groups each with $n$ copies $C_n$-symmetrically aligned around the $y$ axis. Therefore, the polarization and phase indices of the central C line are determined by the minimal value of the phase index of the circle around the $y$ axis: $I_{\text{pl}} = +\frac{I_{\text{ph}}^{\min}}{2}$ with $I_{\text{ph}}^{\min} \in \{I_{\text{ph}}(c) | |I_{\text{ph}}(c)| = \min|2 + nm|, m \in \mathbb{Z}\}$. For $n = 1,2$, the minimal charge is $I_{\text{ph}}^{\min} = 0$, which is consistent with our prediction that there is no C line on the $y$ axis; for $n = 3$, the central C line is first order with $I_{\text{pl}} = \frac{I_{\text{ph}}^{\min}}{2} = -1/2$; for $n = 4$, the central C line is second order with either $I_{\text{pl}} = +1$ or $I_{\text{pl}} = -1$; for $n \geq 5$, the C line must be second order with positive index $I_{\text{pl}} = +1$ (see Supplementary Note 4 for an alternative proof via perturbation analysis). Akin to the cylindrical symmetry case, a left-handed circularly polarized incident wave illuminating a $C_n$ symmetric scatter will induce opposite $I_{\text{ph}}^{\min}$ and **S** but an identical $I_{\text{pl}}$, in comparision to those generated by a right-handed circularly polarized incident wave. This can also be understood from the compatibleness between the line field configurations and the rotational symmetries (see Supplementary Note 4).

As the most interesting case, we generate the central C lines with $I_{\text{pl}} = \pm 1$ protected by the $\bar{C}_4$ rotational symmetry, using the torus nexus with genus $g = 3$ in Fig. 5(c). Under the incidence of the same circularly polarized plane wave, this structure generates complexly distributed C lines. For the ease of visibility, we show the C lines without the torus nexus in Fig. 5(d). A total of twenty C points appear on the surface of the torus nexus: 4 points with index $I_{\text{pl}} = -1$ (marked by the green circles), 8 points with index $I_{\text{pl}} = -\frac{1}{2}$ (corresponding to the tails of the blue lines without a green circle), and 8 points with index $I_{\text{pl}} = +\frac{1}{2}$ (corresponding to the tails of the red lines). The index sum is $\sum I_{pl} = -4$ and agrees with the structure's Euler characteristic $\chi = 2 - 2 \times 3 = -4$. Figure 5(e) shows the polarization ellipses and $\text{Arg}(\Psi)$ on the $y$-mirror plane, where a C point with $I_{ph} = 2I_{pl} = +2$ is clearly observed ($\mathbf{t}_c$ points in $+\hat{\mathbf{y}}$ direction). Figure 5(f) shows the polarization ellipses and $\text{Arg}(\Psi)$ near the green-circled C lines, which clearly shows a C point with index $I_{ph} = 2I_{pl} = -2$.

In the center region of the C lines, as shown by the inset in Fig. 5(d), there are two transition points along $y$ axis where the polarization index changes as $I_{pl} = -1 \rightarrow +1 \rightarrow -1$. At the transition points, four additional first-order C lines (blue) grow out in transverse directions, forming a nexus of C lines. The directions $\mathbf{t}_c$ of the C lines are denoted by the yellow arrows. This remarkable structure of C lines arises from the conservation of the phase index $I_{ph}$. Since $\bar{C}_4$ symmetry supports two different polarization indices $I_{pl} = \pm 1$ of the central C line, the phase index defined on a loop encircling the central C line should be $I_{ph} = 2I_{pl} = \pm 2$, and the sign of the index determines the direction of the C line. When moving the loop along the central C line, the conservation of the phase index forbids the loop to pass through the transition point of two oppositely directed C-line segments (labeled by the large yellow arrows along $y$ axis) without crossing other PSLs, which implies the emergence of the other four transverse C lines connecting with the transition point due to the $\bar{C}_4$ symmetry. The existence of these "budding" C lines is also confirmed via the perturbation analysis near the $y$ axis (see Supplementary Note 4). Still by the conservation of $I_{ph}$ and $\bar{C}_4$ symmetry, the four "budding" C lines are first order and must be directed either all outward or all inward such that the total "arrows" weighted by $|I_{ph}|$ flowing to each nexus point are equal to zero, as shown by the smaller yellow arrows in the inset of Fig. 5(d).

**Discussion**

PSs can emerge in various parameter spaces. In momentum space, they are determined by the Bloch states and are intimately related to the BICs in periodic photonic structures, where V points can give rise to vanishing far-field radiation of Bloch states[53,54]. PSs also provide a direct way to visualize the band Chern number and the topological structure of EPs[32,55]. Such studies typically focus on the PSs that live on the Bloch torus associated with a 2D periodic structure. Extension to other parameters spaces of different topology (genus) is difficult, if not impossible. In the real space, the PSs are attributed to the interference of the Fourier plane waves of the incident and scattering fields. This endows the real-space PSs with several unique properties. They can live on various 2D manifolds and give rise to intriguing PSLs whose configurations can be easily controlled via the topology and symmetry of optical structures, as we have demonstrated here. In addition, the PSs in 3D space naturally carry Pancharatnam-Berry phase and spin-redirection phase[25,56], offering rich mechanisms for manipulating light's polarization and phase. Because of their intrinsic chirality deriving from spin and inhomogeneous phase distribution in the near field, the emergence of the PSs is usually accompanied by the spin-orbit interactions of light[57]. Such spin-orbit interactions can enable directional near-field coupling and far-field radiation with fruitful applications in novel optical sources and nanoantennas[58,59]. Experimentally probing these PSLs can be achieved by using the near-field characterization techniques that can map subwavelength longitudinal and transverse field components, such as scanning nearfield optical microscopy[60].

In sum, we establish a direct relationship between the topology of metal structures and the magnetic PSLs in the near field. We show that the index sum of the PSLs born on the surface of the structures is solely determined by the Euler characteristic of the structures due to the tangent nature of magnetic field near the metal surface. In addition, we find that the interplay of topology with mirror symmetry or generalized rotational symmetry cam give rise to topologically stable higher-order PSs/PSLs lines with richer morphologies including C line nexuses and mirror-symmetric double-twist Möbius strip of polarizations. Remarkably, the topological properties of the PSLs can be well understood by using a non-Hermitian 2-band Hamiltonian, where the topological classification of non-Hermitian EP lines can be mapped to that of C lines characterized by the phase index. According to this correspondence, the merging, bifurcation, and topological transition of C lines in the real space have to observe the phase index conservation law. Our study uncovers exotic topological properties of optical polarization fields that are irrelevant to the specific material or geometry of the optical structures. The results have essentially connected polarization singularities, the topology and symmetry of structures, as well as non-Hermitian physics, opening extra opportunities for fundamental conceptual explorations and many related practical applications in chiral discrimination and sensing, chiral quantum optics, and topological photonics. They may also be extended to other classical wave systems such as sound waves and water-surface waves[61,62].

## Methods

**Numerical simulations**

Full-wave numerical simulations were performed by using a finite-element package COMSOL Multiphysics[63]. The considered structures are made of gold characterized by a Drude model $\varepsilon_{\text{Au}} = 1 - \omega_p^2/(\omega^2 + i\omega\gamma)$, where $\omega_p = 1.28 \times 10^{16}$ rad/s and $\gamma = 7.10 \times 10^{13}$ rad/s[64]. The spheres have radii of 500 nm. The inner and outer radii of the single torus in Fig. 1(c) is $r = 500$ nm and $R = 1500$ nm. For the double-torus in Fig. 1 (d), we set $r = 250$ nm and $R = 750$ nm, and the center distance between the holes is 500 nm. In all cases of Fig. 1, we assume a plane wave propagating in $z$ direction, and the magnetic field is linearly polarized in $y$ direction, and open boundary conditions are applied. In the cases of Fig. 5, we assume that the incident plane wave propagates in $y$ direction and is circularly polarized.

**Homotopy approach characterizing the topology of C lines**

According to the approach for classifying the gapless topological phases of non-Hermitian Hamiltonians[46], the degenerate submanifold $X_D$ (i.e. EPs and non-defective degeneracies) is regarded as the topological obstruction in the configuration space of the Hamiltonians $X_\mathcal{H} = \{\mathcal{H} = \mathbf{H} \cdot \vec{\sigma} : \mathbf{H} \in \mathbb{C}^3\}$, the existence of which induces the nontrivial connectivity to the subspace of $X_\mathcal{H}$ excluding the degeneracies $X = X_\mathcal{H} - X_D = \{\mathcal{H} \in X_\mathcal{H} : \mathbf{H} \cdot \mathbf{H} \neq 0\}$ (i.e. the subspace with separable bands). Tracing a closed loop in $X$, if the loop is noncontractible via continuous deformation, it must encircle the degenerate submanifold in an inextricable manner. Therefore, the topology of the EP lines is essentially encoded by the topological classification of the loops in the nondegenerate subspace $X$, mathematically given by the first homotopy group of $X$. Homotopy refers to a topological equivalence relation. For two loops in $X$ that can be continuously transformed to each other, they belong to the same homotopy equivalence class and hence characterize the same topological phases of degeneracies. And the first homotopy group is just the group formed by all homotopy equivalence classes of loops in $X$. The configuration space of the non-Hermitian Hamiltonian without degeneracies is given by[46] $X = \left(\frac{GL_2(\mathbb{C})}{GL_1(\mathbb{C}) \times GL_1(\mathbb{C})} \times \text{Conf}_2(\mathbb{C})\right)/\mathbb{Z}_2$, where the first part formed by the coset space of the complex general linear groups $GL_n(\mathbb{C})$ represents the space spanned by the two eigenstates of the Hamiltonian, the second part $\text{Conf}_2(\mathbb{C}) = \{\{h_+, h_-\} \in \mathbb{C} : h_+ \neq h_-\}$ denotes the configuration space of two separable complex eigenvalues of $\mathcal{H}(\mathbf{r})$, and the divisor $\mathbb{Z}_2$ denotes the redundancy that the eigenstates and eigenvalues change the orders simultaneously. Moreover, it is shown that the eigenstate and the eigenvalue parts are homotopy equivalent to a sphere and a circle, i.e. $\frac{GL_2(\mathbb{C})}{GL_1(\mathbb{C}) \times GL_1(\mathbb{C})} \simeq S^2$ and $\text{Conf}_2(\mathbb{C}) \simeq S^1$, respectively[46]. Therefore, the configuration space for topological classification can be rewritten as $X \simeq (S^2 \times S^1)/\mathbb{Z}_2$ which can be physically interpreted as the space of the magnetic polarizations (see the main text).

## Acknowledgements

The work described in this paper was supported by a grant from the Research Grants Council of the Hong Kong Special Administrative Region, China (Project No. CityU 11306019). S. W. acknowledge financial support from the National Natural Science Foundation of China (NSFC) (No. 11904306). We thank Prof. C. T. Chan for useful discussions.

## Author contributions

S.W. conceived the idea and supervised the project. J.P. performed the numerical simulations. R.Y.Z. developed the topological classification theory. S.J. and W.L. contributed to the theoretical interpretations. S.W. and R.Y.Z. wrote the draft. All authors contributed to the discussions and the polishing of the manuscript.

## Competing interests

The authors declare no competing interests.

## Reference

1. Hasan, M. Z. & Kane, C. L. Colloquium: Topological insulators. *Rev. Mod. Phys.* **82**, 3045–3067 (2010).

2. Ozawa, T. *et al.* Topological photonics. *Rev. Mod. Phys.* **91**, 015006 (2019).

3. Berry, M. V. & Dennis, M. R. Knotted and linked phase singularities in monochromatic waves. *Proc. R. Soc. Lond. A* **457**, 2251–2263 (2001).

4. Leach, J., Dennis, M. R., Courtial, J. & Padgett, M. J. Knotted threads of darkness. *Nature* **432**, 165–165 (2004).

5. Irvine, W. T. M. & Bouwmeester, D. Linked and knotted beams of light. *Nat. Phys.* **4**, 716–720 (2008).

6. Dennis, M. R., King, R. P., Jack, B., O'Holleran, K. & Padgett, M. J. Isolated optical vortex knots. *Nat. Phys.* **6**, 118–121 (2010).

7. Kedia, H., Bialynicki-Birula, I., Peralta-Salas, D. & Irvine, W. T. M. Tying knots in light fields. *Phys. Rev. Lett.* **111**, 150404 (2013).

8. Bauer, T. *et al.* Observation of optical polarization Möbius strips. *Science* **347**, 964–966 (2015).

9. Bauer, T., Neugebauer, M., Leuchs, G. & Banzer, P. Optical Polarization Möbius Strips and Points of Purely Transverse Spin Density. *Phys. Rev. Lett.* **117**, 013601 (2016).

10. Larocque, H. *et al.* Reconstructing the topology of optical polarization knots. *Nat. Phys.* **14**, 1079 (2018).

11. Pisanty, E. *et al.* Knotting fractional-order knots with the polarization state of light. *Nat. Photon.* **13**, 569–574 (2019).

12. Sugic, D. *et al.* Particle-like topologies in light. *Nat. Commun.* **12**, 6785 (2021).

13. Born, M. & Wolf, E. *Principles of Optics: Electromagnetic Theory of Propagation, Interference and Diffraction of Light.* (Cambridge University Press, Cambridge, 1999).

14. Nye, J. F. & Hajnal, J. V. The wave structure of monochromatic electromagnetic radiation. *Proc. R. Soc. Lond. A* **409**, 21–36 (1987).

15. Schoonover, R. W. & Visser, T. D. Polarization singularities of focused, radially polarized fields. *Opt. Express* **14**, 5733–5745 (2006).

16. Peng, J., Liu, W. & Wang, S. Polarization singularities in light scattering by small particles. *Phys. Rev. A* **103**, 023520 (2021).

17. Grigoriev, K. S., Kuznetsov, N. Yu., Vladimirova, Yu. V. & Makarov, V. A. Fine characteristics of polarization singularities in a three-dimensional electromagnetic field and their properties in the near field of a metallic nanospheroid. *Phys. Rev. A* **98**, 063805 (2018).

18. Yu Kuznetsov, N., Grigoriev, K. S., Vladimirova, Y. V. & Makarov, V. A. Three-dimensional structure of polarization singularities of a light field near a dielectric spherical nanoparticle. *Opt. Express* **28**, 27293 (2020).

19. Chen, W., Yang, Q., Chen, Y. & Liu, W. Global Mie Scattering: Polarization Morphologies and the Underlying Topological Invariant. *ACS Omega* **5**, 14157–14163 (2020).

20. Jia, S., Peng, J., Cheng, Y. & Wang, S. Chiral discrimination by polarization singularities of single metal sphere. *arXiv:2108.00286* (2021).

21. Flossmann, F., O'Holleran, K., Dennis, M. R. & Padgett, M. J. Polarization Singularities in 2D and 3D Speckle Fields. *Phys. Rev. Lett.* **100**, 203902 (2008).

22. Zhang, Y., Yang, X. & Gao, J. Generation of polarization singularities with geometric metasurfaces. *Sci. Rep.* **9**, 19656 (2019).

23. Zhang, Y. *et al.* Observation of Polarization Vortices in Momentum Space. *Phys. Rev. Lett.* **120**, 186103 (2018).

24. Wang, S., Ma, G. & Chan, C. T. Topological transport of sound mediated by spin-redirection geometric phase. *Sci. Adv.* **4**, eaaq1475 (2018).

25. Bliokh, K. Y., Alonso, M. A. & Dennis, M. R. Geometric phases in 2D and 3D polarized fields: geometrical, dynamical, and topological aspects. *Rep. Prog. Phys.* **82**, 122401 (2019).

26. Zhen, B., Hsu, C. W., Lu, L., Stone, A. D. & Soljačić, M. Topological nature of optical bound states in the continuum. *Phys. Rev. Lett.* **113**, 257401 (2014).

27. Doeleman, H. M., Monticone, F., den Hollander, W., Alù, A. & Koenderink, A. F. Experimental observation of a polarization vortex at an optical bound state in the continuum. *Nat. Photon.* **12**, 397–401 (2018).

28. Liu, W. *et al.* Circularly Polarized States Spawning from Bound States in the Continuum. *Phys. Rev. Lett.* **123**, 116104 (2019).

29. Fang, C., Weng, H., Dai, X. & Fang, Z. Topological nodal line semimetals. *Chin. Phys. B* **25**, 117106 (2016).

30. Armitage, N. P., Mele, E. J. & Vishwanath, A. Weyl and Dirac semimetals in three-dimensional solids. *Rev. Mod. Phys.* **90**, 015001 (2018).

31. Chen, W., Yang, Q., Chen, Y. & Liu, W. Evolution and global charge conservation for polarization singularities emerging from non-Hermitian degeneracies. *Proc. Natl. Acad. Sci. USA* **118**, (2021).

32. Zhou, H. *et al.* Observation of bulk Fermi arc and polarization half charge from paired exceptional points. *Science* **359**, 1009–1012 (2018).

33. Shen, H., Zhen, B. & Fu, L. Topological Band Theory for Non-Hermitian Hamiltonians. *Phys. Rev. Lett.* **120**, 146402 (2018).

34. Yang, Z., Schnyder, A. P., Hu, J. & Chiu, C.-K. Fermion Doubling Theorems in Two-Dimensional Non-Hermitian Systems for Fermi Points and Exceptional Points. *Phys. Rev. Lett.* **126**, 086401 (2021).

35. Wang, K., Dutt, A., Wojcik, C. C. & Fan, S. Topological complex-energy braiding of non-Hermitian bands. *Nature* **598**, 59–64 (2021).

36. Biagioni, P., Huang, J.-S. & Hecht, B. Nanoantennas for visible and infrared radiation. *Rep. Prog. Phys.* **75**, 024402 (2012).

37. Kadic, M., Milton, G. W., van Hecke, M. & Wegener, M. 3D metamaterials. *Nat. Rev. Phys.* **1**, 198–210 (2019).

38. Chen, H.-T., Taylor, A. J. & Yu, N. A review of metasurfaces: physics and applications. *Rep. Prog. Phys.* **79**, 076401 (2016).

39. Freund, I. Optical Möbius strips, twisted ribbons, and the index theorem. *Opt. Lett.* **36**, 4506 (2011).

40. Freund, I. Polarization Möbius strips on elliptical paths in three-dimensional optical fields. *Opt. Lett.* **45**, 3333 (2020).

41. Kuznetsov, N. Yu., Grigoriev, K. S. & Makarov, V. A. Topology of polarization-ellipse strips in the light scattered by a dielectric nanosphere. *Phys. Rev. A* **104**, 043505 (2021).

42. Berry, M. V. Index formulae for singular lines of polarization. *J. Opt. A: Pure Appl. Opt.* **6**, 675–678 (2004).

43. Berry, M. V. Much ado about nothing: optical distortion lines (phase singularities, zeros, and vortices). in *International Conference on Singular Optics* vol. 3487 1–5 (SPIE, 1998).

44. Jackson, J. D. *Classical Electrodynamics*. (Wiley, New York, 1998).

45. Needham, T. *Visual Differential Geometry and Forms: A Mathematical Drama in Five Acts*. (Princeton University Press, Princeton, 2021).

46. Wojcik, C. C., Sun, X.-Q., Bzdušek, T. & Fan, S. Homotopy characterization of non-Hermitian Hamiltonians. *Phys. Rev. B* **101**, 205417 (2020).

47. Freund, I. Optical Möbius strips in three-dimensional ellipse fields: I. Lines of circular polarization. *Opt. Commun.* **283**, 1–15 (2010).

48. Freund, I. Multitwist optical Möbius strips. *Opt. Lett.* **35**, 148–150 (2010).

49. Galvez, E. J. *et al.* Multitwist Möbius Strips and Twisted Ribbons in the Polarization of Paraxial Light Beams. *Sci. Rep.* **7**, 13653 (2017).

50. Berry, M. V., Dennis, M. R. & Lee, R. L. Polarization singularities in the clear sky. *New J. Phys.* **6**, 162–162 (2004).

51. Feynman, R. P., Leighton, R. B. & Sands, M. *The Feynman Lectures on Physics, Vol. III: Quantum Mechanics*. (Basic Books, London, 2011).

52. Yang, Q., Chen, W., Chen, Y. & Liu, W. Symmetry Protected Invariant Scattering Properties for Incident Plane Waves of Arbitrary Polarizations. *Laser Photonics Rev.* **15**, 2000496 (2021).

53. Ye, W., Gao, Y. & Liu, J. Singular Points of Polarizations in the Momentum Space of Photonic Crystal Slabs. *Phys. Rev. Lett.* **124**, 153904 (2020).

54. Hsu, C. W. *et al.* Observation of trapped light within the radiation continuum. *Nature* **499**, 188–191 (2013).

55. Fösel, T., Peano, V. & Marquardt, F. L lines, C points and Chern numbers: understanding band structure topology using polarization fields. *New J. Phys.* **19**, 115013 (2017).

56. Berry, M. V. The Adiabatic Phase and Pancharatnam's Phase for Polarized Light. *J. Mod. Opt.* **34**, 1401–1407 (1987).

57. Bliokh, K. Y., Rodríguez-Fortuño, F. J., Nori, F. & Zayats, A. V. Spin-orbit interactions of light. *Nat. Photon.* **9**, 796–808 (2015).

58. Wang, S. B. & Chan, C. T. Lateral optical force on chiral particles near a surface. *Nat. Commun.* **5**, 3307 (2014).

59. Wang, S. *et al.* Arbitrary order exceptional point induced by photonic spin–orbit interaction in coupled resonators. *Nat. Commun.* **10**, 832 (2019).

60. Rotenberg, N. & Kuipers, L. Mapping nanoscale light fields. *Nat. Photon.* **8**, 919–926 (2014).

61. Wang, S. *et al.* Spin-orbit interactions of transverse sound. *Nat. Commun.* **12**, 6125 (2021).

62. Bliokh, K. Y. *et al.* Polarization singularities and Möbius strips in sound and water-surface waves. *Phys. Fluids* **33**, 077122 (2021).

63. www.comsol.com.

64. Olmon, R. L. *et al.* Optical dielectric function of gold. *Phys. Rev. B* **86**, (2012).

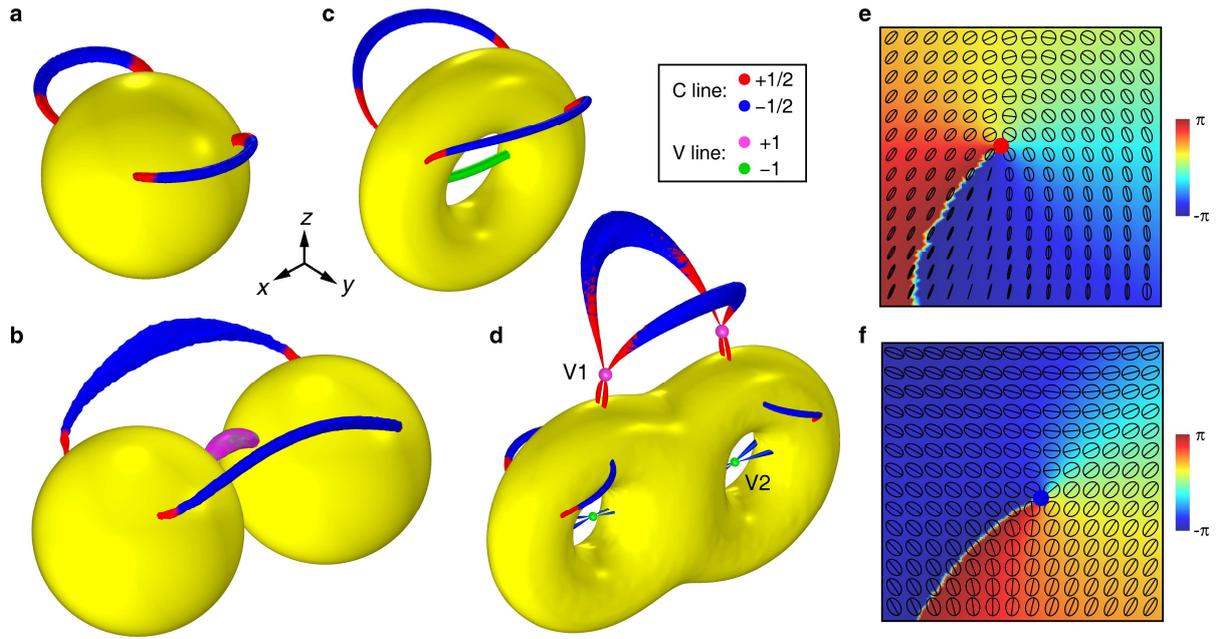

**Fig. 1 Polarization singularities generated by topological structures.** (a)-(d) C lines (blue/red) and V lines (magenta/green) generated by structures with Euler characteristic $\chi = +2, +4, 0, -2$. Polarization ellipses and $\mathrm{Arg}(\Psi)$ near the C points with $I_{\mathrm{pl}} = +\frac{1}{2}$ (e) and $I_{\mathrm{pl}} = -\frac{1}{2}$ (f).

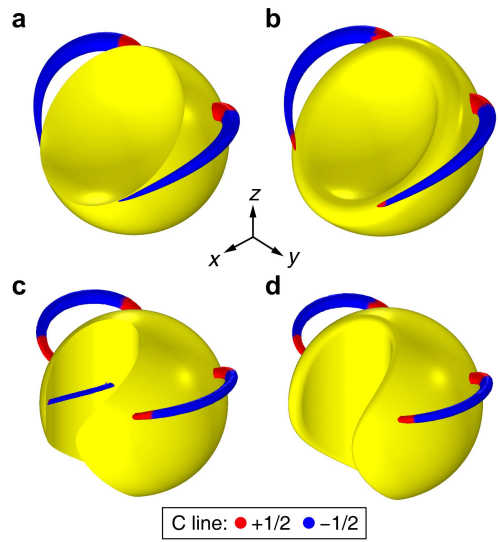

**Fig. 2 Effect of singular edges.** C lines generated by a sphere with singular edges ((a), (c)) and with edge singularities smoothed out ((b), (d)). There are two C lines in (a), (b) and (d), and six C lines in (c).

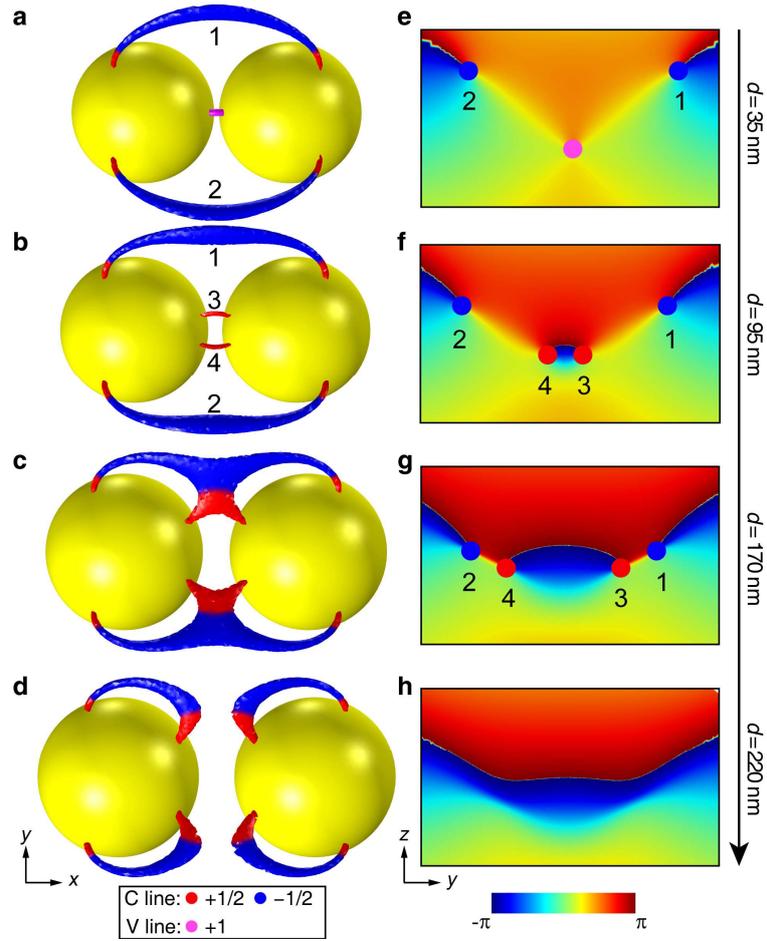

**Fig. 3 Topological transition of PSLs.** (a)-(d) Topological transition of PSLs when two spheres are gradually separated. The accidental V line in the gap bifurcates into two C lines (labeled as 3 and 4) which further merge with rest C lines (labeled as 1 and 2) to enable the topological transition. (e)-(h) Arg(Ψ) on the *yz*-mirror plane at different separations of the spheres corresponding to (a)-(d).

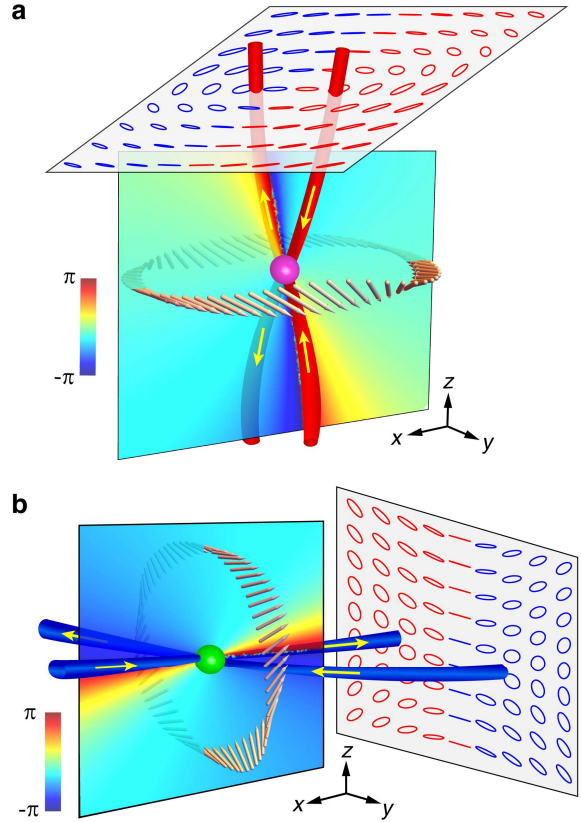

**Fig. 4 V points due to mirror symmetry.** (a) V1 point due to the crossing of two C lines with $I_{pl} = +\frac{1}{2}$. (b) V2 point due to crossing of two C lines with $I_{pl} = -\frac{1}{2}$. The blue (red) color of the polarization ellipses corresponds to negative (positive) spin. The colors on the *xz*-mirror plane at the V points show $\text{Arg}(\Psi)$. The pink arrows in (a) and (b) denote the polarization major axis **A** on a self-mirror symmetric loop, showing a double-twist Möbius strip. The yellow arrows show the directions of the C lines.

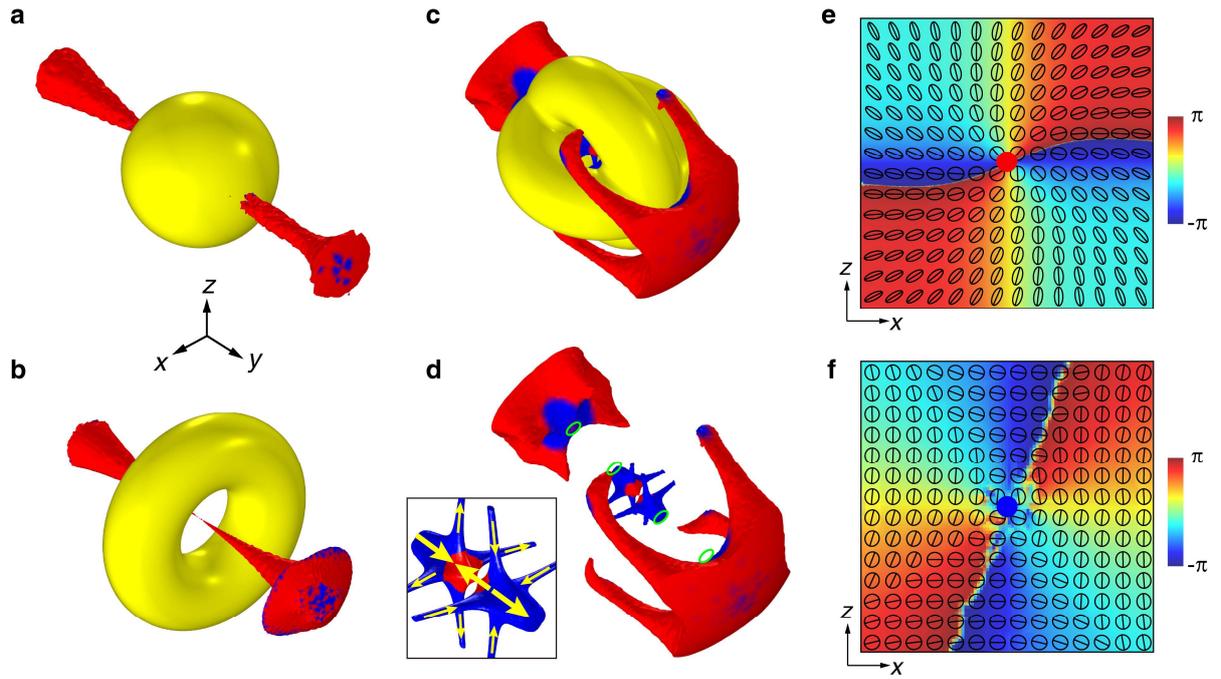

**Fig. 5 Higher-order C lines due to the generalized rotational symmetry.** Higher-order C lines with index $I_{pl} = +1$ generated by (a) a sphere and (b) a torus. (c),(d) Higher-order C lines generated by a torus nexus with C$_4$ rotational symmetry. The inset shows a zoom-in of the center C lines, where the yellow arrows show the directions of the C lines. Polarization ellipses and $\mathrm{Arg}(\Psi)$ for the C points in (c) with indices $I_{pl} = +1$ (e) and $I_{pl} = -1$ (f).

Supplementary Information for

# Topological near fields generated by topological structures

Jie Peng[1,*], Ruo-Yang Zhang[2,*], Shiqi Jia[1], Wei Liu[3], and Shubo Wang[1,4,†]

[1]*Department of Physics, City University of Hong Kong, Tat Chee Avenue, Kowloon, Hong Kong, China*
[2]*Department of Physics, The Hong Kong University of Science and Technology, Clear Water Bay, Kowloon, Hong Kong, China*
[3]*College for Advanced Interdisciplinary Studies, National University of Defense Technology, Changsha, Hunan 410073, China*
[4]*City University of Hong Kong Shenzhen Research Institute, Shenzhen, Guangdong 518057, China*

\* These authors contributed equally.
† Correspondence should be addressed to: Shubo Wang (shubwang@cityu.edu.hk)

## Supplementary Note 1: Topological mapping of the near field

Under the excitation of an incident wave, currents will be induced in the metal structures. If the skin depth is much smaller compared to the dimension of the structures, the induced currents localize on a thin surface layer. In this case, the magnetic field **H** near the surface is dominated by the tangent component $H_\parallel$ [1]. Figure S1 shows the numerical simulation results of $\frac{H_\perp}{|\mathbf{H}|}$ for structures with different genuses (i.e., single sphere, double spheres, torus, double-torus, and torus nexus) and within the frequency range of interest. We see that the normal components $H_\perp$ of the magnetic field are indeed very small. Thus, the corresponding polarization ellipses of the magnetic field are parallel to the surface. The bivector major axis **A** can be mapped to a line field defined on a 2D smooth manifold (i.e., the surface of the structures). Applying the Poincaré-Hopf theorem to this line field, we then can predict the total topological index of the singularities (i.e., C points and V points).

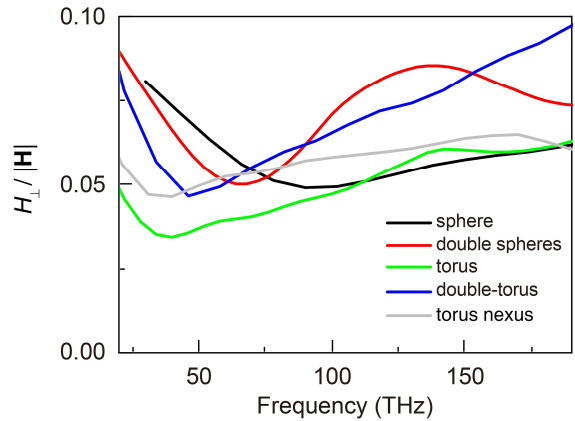

**Fig. S1.** Normal component of the total magnetic field near the surface of metal structures with different genuses.

## Supplementary Note 2: Relation between polarization index and phase index

In previous studies, two different ways to define the topological indices of a C line in 3D space, i.e., the definitions with the polarization ellipse and with the phase winding number, were proposed and their intrinsic relation had been explored [2–5]. Here, we review these two kinds of topological indices from the perspective of topological classification.

### 1. Local and global polarization indices

As shown in Fig. S2(a), consider a C point, $\mathbf{r}_c$, with circular polarization $\mathbf{H}_c = \frac{1}{\sqrt{2}} H_c(\mathbf{e}_1 + i\mathbf{e}_2)$, where $\mathbf{e}_1$ and $\mathbf{e}_2$ are orthogonal unit vectors in the local polarization plane of $\mathbf{H}_c$. The magnetic spins are approximately aligned along the same direction $\mathbf{S}_c = \mathbf{S}(\mathbf{r}_c) = 2H_c^2 \mathbf{e}_1 \times \mathbf{e}_2$, namely the polarization ellipses are approximately in the same plane perpendicular to $\mathbf{e}_s = \frac{\mathbf{S}_c}{2H_c^2}$, characterized by the major axis $\mathbf{A} = A_1 \mathbf{e}_1 + A_2 \mathbf{e}_2 + O(\mathbf{r} - \mathbf{r}_c)\mathbf{e}_s$. As such, we can assign a *local polarization index* to the C point as the winding number of the major axis in the local polarization plane

$$
\begin{aligned}
I_{\text{pl}}(\mathbf{r}_c) &= \lim_{\gamma \to \mathbf{r}_c} \frac{1}{2\pi} \oint_{\gamma_c} d\arctan\left(\frac{A_2}{A_1}\right) = \lim_{\gamma \to \mathbf{r}_c} \frac{1}{2\pi} \oint_\gamma \frac{1}{A^2} (\mathbf{A} \times d\mathbf{A}) \cdot \mathbf{e}_s \\
&= \lim_{\gamma \to \mathbf{r}_c} \frac{1}{2\pi} \oint_\gamma \frac{1}{AB} (\mathbf{B} \cdot d\mathbf{A}) = \lim_{\gamma \to \mathbf{r}_c} \frac{1}{2\pi} \oint_\gamma \frac{2}{H_c^2} (\mathbf{B} \cdot d\mathbf{A})
\end{aligned}
\tag{S1}
$$

where the integration is along an infinitesimal circle $\gamma$ around $\mathbf{r}_c$ with positive normal vector $\mathbf{S}_c$. If we map the transverse polarization ellipses near $\mathbf{r}_c$ to the Poincaré sphere, the azimuthal angle on the sphere is given by $\phi = 2\arctan(A_2/A_1)$. Thus, the local polarization index can be equivalently expressed as $I_{\text{pl}} = \lim_{\gamma \to \mathbf{r}_c} \frac{1}{4\pi} \oint_\gamma d\phi$. We note that all the polarization indices mentioned in the main text refer to this local polarization index, unless otherwise stated. Equation (S1) shows that $I_{\text{pl}}$ is just given by the geometric phase integrated by the Euler connection (differential 1-form) $a_{\text{Eu}} = \frac{1}{AB} (\mathbf{B} \cdot d\mathbf{A}) = \mathbf{e}_B \cdot d\mathbf{e}_A$ of the orthonormal frame $(\mathbf{e}_A, \mathbf{e}_B) = \left(\frac{\mathbf{A}}{A}, \frac{\mathbf{B}}{B}\right)$. And $\lim_{\mathbf{r} \to \mathbf{r}_c} A = \lim_{\mathbf{r} \to \mathbf{r}_c} B = \frac{1}{\sqrt{2}} H_c$ has been used in the last step. In the absence of symmetry protection, only first order C lines can stably exist in the space, around which the major axes form a half vortex, hence the local polarization index of a first-order C line is given by $I_{\text{pl}} = \pm\frac{1}{2}$. However, there may exit some singular points on a C line at which $\mathbf{S}_c \cdot \mathbf{t} = 0$, and one cannot find a circle $\gamma$ that is both normal to $\mathbf{S}_c$ and encircles the C line at such a singular point. As a result, the sign of $I_{\text{pl}}$ can flip at such points, and hence the local polarization index $I_{\text{pl}}$ is not a conserved quantity along a C line. Such sign flips of $I_{\text{pl}}$ are very common and occur in almost all the cases considered in the main text. As an additional note, the local polarization index of a V point in 3D polarization fields is indeed ill-defined, since in general there does not exist a special direction, as the direction of **S** for a C point, that tends to be perpendicular to all polarization ellipses near the V point. Nevertheless, the *ad hoc* polarization indices of the accidental V lines shown in Fig. 1 is suitable for the application of Poincaré-Hopf index theorem. This is because the magnetic fields are perpendicular to the surface normal on the structural surfaces which forms a fixed rotation axis of the tangent polarization ellipses around the V point on the surface.

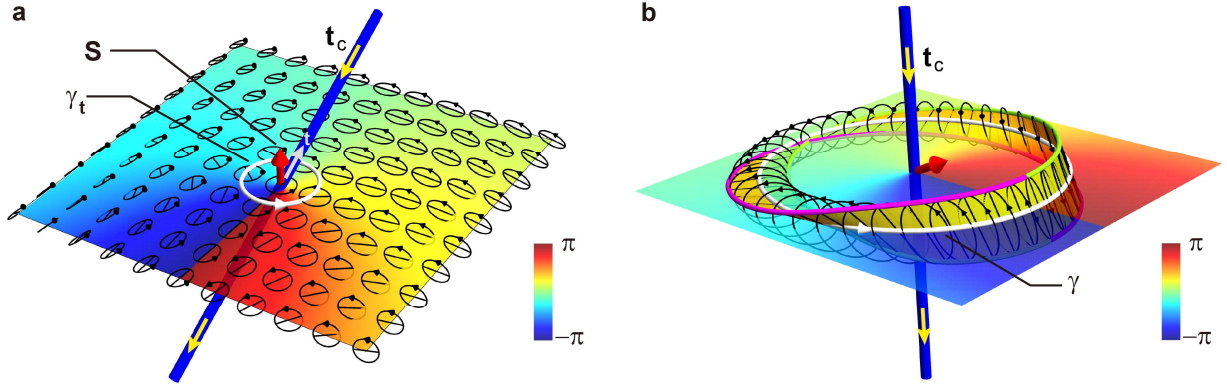

**Fig. S2.** Schematic of the relations between local polarization index, global polarization index, and phase index around a C line (blue). (a) Phase vortex and polarization ellipses lying on the plane perpendicular to the magnetic spin **S** (red arrow) of the central C point on a C line. (b) The Möbius strip of polarization ellipses along a finite loop $\gamma$ (white circle) around a C line.

On the other hand, the definition of the local polarization index in Eq. (S1) cannot be used to define the index of 3D polarizations along a loop with finite size, since the integral $\oint a_{\text{Eu}}$ is not quantized in general. If we only concern about the major axis **A** of the polarization ellipse, the globally conserved polarization index along an arbitrary loop, $\gamma$, can be introduced by the homotopy equivalence class of loops in the real projective space $\mathbb{R}P^2$ (the space of **A**) [6]:

$$\nu_{\text{pl}}(\gamma) = 0 \text{ or } 1 \in \pi_1(\mathbb{R}P^2) = \mathbb{Z}_2. \tag{S2}$$

This shows that the alignments of the major axis along a loop $\gamma$ can only be classified into two topological phases (homotopy classes), i.e., the trivial phase ($\nu_{\text{pl}}(\gamma) = 0$) corresponding to the untwisted alignment of the major axes and the nontrivial phase ($\nu_{\text{pl}}(\gamma) = 1$) corresponding to the Möbius twisted alignment (see Fig. S2(b)). Note that here untwisted alignment and Möbius twisted alignment should be understood in a topologically stable sense. Although Multi-twisted Möbius strips of polarization had been widely studied in literature [7–10], the twist number $n\ (\geq 2)$ is not a topological invariant but can change under the continuous deformation of the loop or of the optical fields (this is because the polarization ellipses are not solid stuffs, and the self-intersection of the polarization strip is allowed). Only the parity of the twist number, i.e., $v_{\text{pl}}(\gamma)$, is topologically stable against the continuous deformation [3].

## 2. Phase index

A general monochromatic magnetic field can be expressed as $\mathbf{H} = (\mathbf{A} + i\mathbf{B})e^{i\theta}$, where $\theta = \arg(\mathbf{H} \cdot \mathbf{H})/2$ is a proper phase. As we have shown in the main text, by introducing the scalar field $\Psi = \mathbf{H} \cdot \mathbf{H} = H^2 e^{i2\theta}$, the phase winding number of the scalar field $\Psi$ along a loop $\gamma$, $I_{\text{ph}}(\gamma) = \frac{1}{2\pi}\oint_\gamma d\arg(\Psi) = \frac{1}{\pi}\oint_\gamma d\theta \in \mathbb{Z}$, gives a complete topological classification of the C points enclosed by 1D loops [11]. Here, we consider the relation between the phase index $I_{\text{ph}}$, the global polarization index $v_{\text{pl}}$, and the local polarization index $I_{\text{pl}}$.

First, we consider a loop $\gamma$ with finite size. Starting at a base point $\mathbf{r}_0$ on the loop, we examine the continuous evolution of the major-axis bivector and the phase, i.e., $(\mathbf{A}, e^{i\theta})$, along the loop, which begin as $(\mathbf{A}_0, e^{i\theta_0})$ at the base point $\mathbf{r}_0$ and finally turn to $(\mathbf{A}_f, e^{i\theta_f})$ when returning to $\mathbf{r}_0$. Since $\mathbf{A}e^{i\theta}(\mathbf{r}_0) = \mathbf{A}_0 e^{i\theta_0} = \mathbf{A}_f e^{i\theta_f}$, we have either $(\mathbf{A}_f, e^{i\theta_f}) = (\mathbf{A}_0, e^{i\theta_0})$ or $(\mathbf{A}_f, e^{i\theta_f}) = (-\mathbf{A}_0, -e^{i\theta_0})$. Thus, we obtain the relation between the phase index and the global polarization index along the loop $\gamma$:

$$\exp[i\pi I_{\text{ph}}(\gamma)] = \exp\left[i\oint_\gamma d\theta\right] = \exp[i(\theta_f - \theta_0)] = \text{sign}(\mathbf{A}_0 \cdot \mathbf{A}_f) = \exp[i\pi v_{\text{pl}}(\gamma)] \quad (S3)$$

or equivalently,

$$v_{\text{pl}}(\gamma) = I_{\text{ph}}(\gamma) \bmod 2. \quad (S4)$$

As a result, we know that if the major axes form a trivial untwisted (Möbius twisted) strip along a loop, the loop encloses even (odd) numbers of first-order C lines. Hence, although both the phase index $I_{\text{ph}}(\gamma)$ and the global polarization index $v_{\text{pl}}(\gamma)$ along a loop are invariant against continuous deformations, only the phase index can distinguish all inequivalent topological phases associated with C lines enclosed by the loop.

In contrast to the phase index $I_{\mathrm{ph}}(\gamma)$ and the global polarization index $\nu_{\mathrm{pl}}(\gamma)$ that are generally defined for a loop $\gamma$ in the 3D space, the local polarization index $I_{\mathrm{pl}}(\mathbf{r}_c)$ characterizes the property of a certain C point $\mathbf{r}_c$. Even so, if a loop $\gamma_t$ only encircles a single C line as shown in Fig. S2(a), the corresponding phase index $I_{\mathrm{ph}}(\gamma_t)$ serves as a minimal charge of the C line and has a definite relation to the local polarization index $I_{\mathrm{pl}}$ at each point on the C line. To derive this relation, we introduce the complex polarization vector $\mathbf{C} = \mathbf{A} + i\mathbf{B} = \mathbf{H}e^{-i\theta}$, the Berry connection of the complex vector $\mathbf{C}$ is identical with the Euler connection of the real vector frame

$$a_{\mathrm{Berry}}(\mathbf{C}) = \frac{i\mathbf{C}^* \cdot d\mathbf{C}}{|\mathbf{C}|^2} = \frac{\mathbf{B}\cdot d\mathbf{A} - \mathbf{A}\cdot d\mathbf{B}}{H^2} = \frac{2}{H^2}(\mathbf{B}\cdot d\mathbf{A}). \tag{S5}$$

where $d(\mathbf{B}\cdot\mathbf{A}) = \mathbf{B}\cdot d\mathbf{A} + \mathbf{A}\cdot d\mathbf{B} = 0$ has been used. Therefore, the local polarization index of a C point $\mathbf{r}_c$ can be expressed as

$$I_{\mathrm{pl}}(\mathbf{r}_c) = \lim_{\gamma_t \to \mathbf{r}_c} \frac{1}{2\pi}\oint_{\gamma_t} a_{\mathrm{Eu}} = \lim_{\gamma_t \to \mathbf{r}_c} \frac{1}{2\pi}\oint_{\gamma_t} a_{\mathrm{Berry}}(\mathbf{C}). \tag{S6}$$

Further, according to $\mathbf{C} = \mathbf{H}e^{-i\theta}$, we have

$$a_{\mathrm{Berry}}(\mathbf{C}) = \frac{i\mathbf{C}^*\cdot d\mathbf{C}}{|\mathbf{C}|^2} = \frac{i\mathbf{H}^*\cdot d\mathbf{H}}{|\mathbf{H}|^2} + d\theta = a_{\mathrm{Berry}}(\mathbf{H}) + \frac{1}{2}d\arg(\Psi). \tag{S7}$$

Since $\mathbf{H}$ field is continuous at $\mathbf{r}_c$ with no singularity, the integration of $a_{\mathrm{Berry}}(\mathbf{H})$ along the infinitesimal circle must vanish, and hence we can obtain the relation between the local polarization index of a C point and the phase index along an infinitesimal circle $\gamma_t$ around the C point, provided that $\mathbf{S}_c\cdot\mathbf{t} \neq 0$ ($\mathbf{t}$ is a tangent vector of the C line with the direction complying with the right-hand rule of the loop $\gamma_t$) [4,5]:

$$I_{\mathrm{pl}}(\mathbf{r}_c) = \mathrm{sign}(\mathbf{S}_c\cdot\mathbf{t})\left[\lim_{\gamma_t\to\mathbf{r}_c}\frac{1}{2\pi}\oint_{\gamma_t} a_{\mathrm{Berry}}(\mathbf{H}) + \frac{1}{4\pi}\oint_{\gamma_t} d\arg(\Psi)\right] \tag{S8}$$

$$= \mathrm{sign}(\mathbf{S}_c\cdot\mathbf{t})\frac{1}{4\pi}\oint_{\gamma_t} d\arg(\Psi) = \frac{1}{2}\mathrm{sign}(\mathbf{S}_c\cdot\mathbf{t})I_{\mathrm{ph}}(\gamma_t).$$

In terms of the phase index along a loop $\gamma_t$ encircling the C line, we can uniquely define a positive direction to the C-line by the directed tangent vector (see Fig. S2(a))

$$\mathbf{t}_c = \mathrm{sign}\left(I_{\mathrm{ph}}(\gamma_t)\right)\mathbf{t} = \mathrm{sign}(I_{\mathrm{pl}}\,\mathbf{S}_c\cdot\mathbf{t})\mathbf{t}. \tag{S9}$$

# Supplementary Note 3: Mirror symmetry protected $\mathbb{Z}_2$ topology of polarization on mirror-symmetric loops

## 1. Topology along semi-loop terminated on mirror plane – relative homotopy approach

Since the system respects mirror symmetry, the distributions of the magnetic fields on the two sides of the mirror plane $\Pi$ are one-to-one correspondent. Hence, to characterize the topology along a self-mirror-symmetric loop $c$, one only need to investigate a semi-loop of $c$ on either side of $\Pi$ (see Fig. S3(a)). For such a semi-loop $\gamma$ (which is equivalent to a 1D disk $D^1$) terminated at two points $\mathbf{r}_0$, $\mathbf{r}_t$ ($\in \Pi$) on the mirror-plane and supposing $\mathbf{r}_0 \in \Pi$ is fixed, while $\mathbf{r}_t$ can move freely in $\Pi$, the magnetic field $\mathbf{H}$ along $\gamma$ defines a map (see Fig. S3(a)) [6,12]:

$$\mathbf{H} \circ \gamma: \begin{cases} D^1 \to X \simeq \dfrac{S^2 \times S^1}{\mathbb{Z}_2} \\ \partial D^1 \to X_\Pi \simeq S^1 \\ \partial D^1 \ni \mathbf{r}_0 \to \mathbf{H}_0 \in X_\Pi \end{cases}. \tag{S10}$$

The first line of the map means that the magnetic fields of all points on $\gamma$ belong to the configuration space $X$; the second line means that the two terminal points of $\gamma$ belong to $X_\Pi$; the third line indicates that the magnetic field at the starting point of $\gamma$ is fixed to the base point $\mathbf{H}_0 \in X_\Pi$. The topological equivalence of $\mathbf{H}(\mathbf{r})$ along two semi-loops $\gamma_1$, $\gamma_2$ requires that the distributions of the magnetic fields on the two paths can continuously deform to each other without encountering a singular point (i.e. a C point or V point), which mathematically means $\mathbf{H} \circ \gamma_1 \sim_{X_\Pi} \mathbf{H} \circ \gamma_2$ ($\sim_{X_\Pi}$ denotes the homotopy equivalence relative to $X_\Pi$), namely there exists a homotopy (i.e. a continuous deformation between $\mathbf{H} \circ \gamma_1$ and $\mathbf{H} \circ \gamma_2$) $F: D^1 \times [0,1] \to X$ relative to $X_\Pi$ such that:

$$F(\cdot,0) = \mathbf{H} \circ \gamma_1, \quad F(\cdot,1) = \mathbf{H} \circ \gamma_2, \quad F(\partial D^1, t) \subset X_\Pi, \quad F(\mathbf{r}_0, t) \equiv \mathbf{H}_0, \tag{S11}$$

where the first (second) equality denotes that the function $F(\cdot, t)$ is equal to the map along $\gamma_1$ ($\gamma_2$) at $t = 0$ ($t = 1$), the third and last equalities indicate that the values of $F$ at the terminals of the semi-loop are restricted in $X_\Pi$ and the base point is fixed at $\mathbf{H}_0$ during the deformation. Then we know that the topological classification of the magnetic fields along a semi-loop $\gamma$ (or a self-mirror-symmetric loop) is determined by the *relative homotopy group*

$$\pi_1(X, X_\Pi, \mathbf{r}_0) = \pi_1\left(\dfrac{S^2 \times S^1}{\mathbb{Z}_2}, S^1, \mathbf{r}_0\right) = \{\mathbf{H} \circ \gamma\}/\sim_{X_\Pi}. \tag{S12}$$

Since $X_\Pi \simeq S^1$ is path connected, $\pi_1(X, X_\Pi, \mathbf{r}_0) = \pi_1(X, X_{\Pi_1}, \mathbf{r}'_0)$ for any two different base points $\mathbf{r}_0$, $\mathbf{r}'_0 \in \Pi$, so we may simply express the relative homotopy group as $\pi_1(X, X_\Pi)$. To compute the relative homotopy group $\pi_1(X, X_\Pi)$, we consider the natural fibration of the configuration space $X$, which is defined by the projection map from the configuration space $X$ to the space of the major axis $\mathbb{R}P^2$ [6,11]:

$$p: X \simeq \frac{S^2 \times S^1}{\mathbb{Z}_2} \quad \rightarrow \quad \mathbb{R}P^2 = \{\boldsymbol{A} \in S^2 | \boldsymbol{A} \sim -\boldsymbol{A}\} \tag{S13}$$

such that $p(\mathbf{H}) = [\pm \boldsymbol{A}]$ gives the bivector of the major axis of the polarization ellipse for each point $\mathbf{H} \in X$, and the fiber at each point $[\pm \boldsymbol{A}] \in \mathbb{R}P^2$ : $p^{-1}([\pm \boldsymbol{A}]) = \{\mathbf{H} = \boldsymbol{A} e^{i\phi} | \phi \in [0, 2\pi]\} \simeq S^1 \simeq U(1)$. Therefore, *the projection $p: X \rightarrow \mathbb{R}P^2$ defines a $U(1)$-principle bundle on the base space $\mathbb{R}P^2$* (i.e. the "polarization sphere" of the major axis, as shown in Fig. S3(d) where any two antipodal points on the sphere are identified as the same point in $\mathbb{R}P^2$).

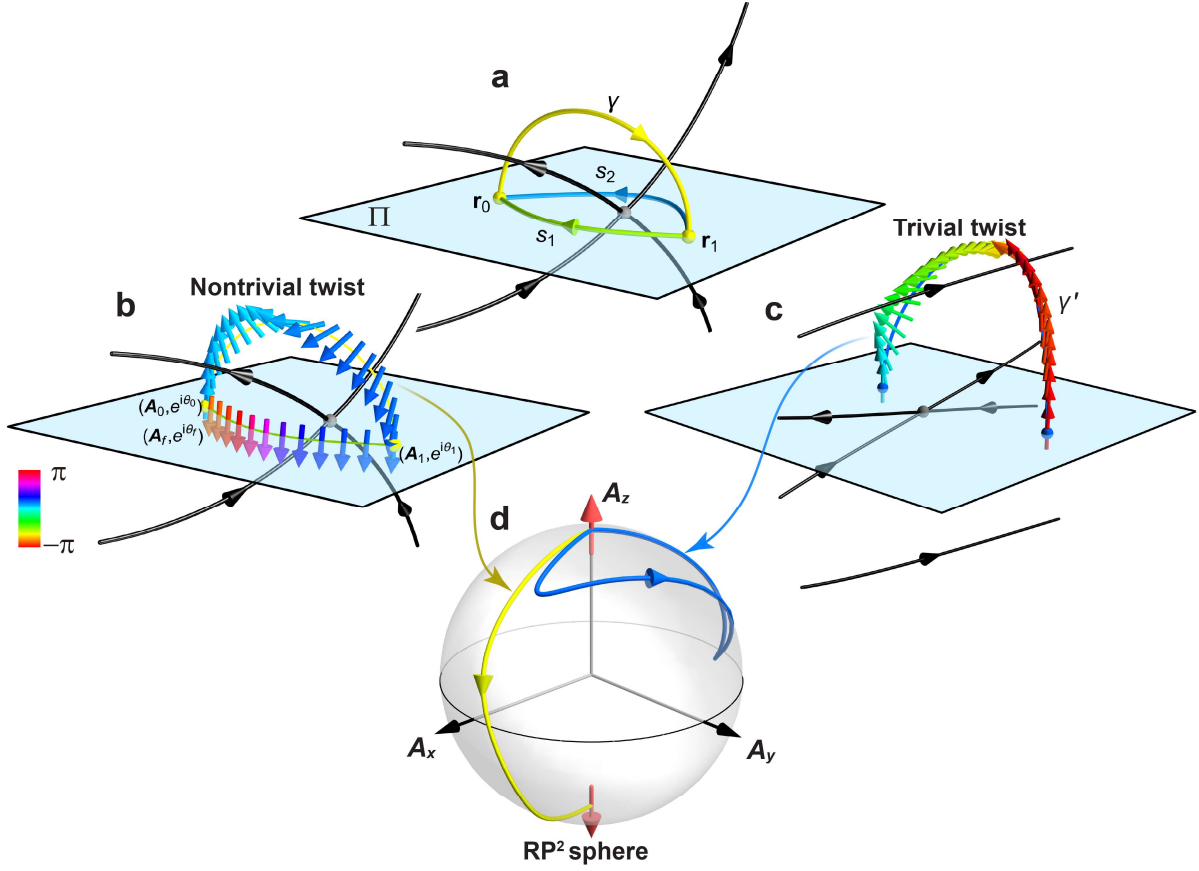

**Fig. S3.** $\mathbb{Z}_2$ topology on a semi-loop $\gamma$ terminated on the mirror plane $\Pi$. The back lines in (a-c) denote mirror-partner C lines and their crossings are V points in $\Pi$, where the arrows on the C lines shown their positive directions. The colored arrows in (b,c) denote the vectors **A** of major axis along the loop $\gamma_1 = \gamma \circ s_1$ (b) and $\gamma'$ (c), and the color of the arrows displays the phase angle $\theta$. (d) Nontrivial and trivial trajectories on the $\mathbb{R}P^2$ sphere corresponding to the Mobius and trivial twists of the major axis along the semi-loops $\gamma$ and $\gamma'$ in (b) and (c), respectively.

According to Theorem 4.41 in Ref. [6] for a fiber bundle $p: E \to B$ that the $n$-homotopy group of the bundle $E$ relative to the fiber $F$ is isomorphic to the $n$-homotopy group of the base manifold by the induced map $p_*: \pi_n(E, F, x_0) \to \pi_n(B, b_0)$ with $x_0 \in p^{-1}(b_0)$, we obtain the topological classification on the semi-loops protected by $y$-mirror symmetry

$$\pi_1(X, X_{\Pi_1}) = \pi_1\left(\frac{S^2 \times S^1}{\mathbb{Z}_2}, S^1\right) \stackrel{p_*}{=} \pi_1(\mathbb{R}P^2) = \mathbb{Z}_2. \tag{S14}$$

Recalling that $\mathbb{R}P^2$ is just the space of the major axis, Eq. (S14) reveals that the topology along semi-loops terminated in the mirror plane $\Pi$ is absolutely determined by the twist of the major axes along the semi-loops, which are classified into two phases characterized by the two kinds of homotopic inequivalent trajectories on the $\mathbb{R}P^2$ sphere. Since the major axes at the two terminal points of a semi-loop is fixed along the normal direction of $\Pi$ corresponding to the south and north poles of the $\mathbb{R}P^2$ sphere, the trajectories of **A** on the $\mathbb{R}P^2$ sphere should either form arcs connecting the two poles (these arcs is already closed in $\mathbb{R}P^2$), or form loops that start and end at the same pole. For the arcs on the $\mathbb{R}P^2$ sphere, the major axis exhibits an inextricable odd twist along the semi-loops (see Fig. S3(b)), thereby deemed as topologically nontrivial phase. In contrast, since the loops on the $\mathbb{R}P^2$ sphere are contractible to a point, the major axes **A** along the semi-loop exhibit a trivial twist (see Fig. S3(c)) and can be continuously deformed into the configuration that all **A** are parallelly aligned. Equivalently, along a self-mirror-symmetric loop $c$ (the corresponding semi-loop is $\gamma$), the major axes in the trivial and nontrivial phases form an untwisted strip and a mirror-symmetric double-twisted Möbius strip, respectively, which can be characterized by the global polarization index

$$v_{\text{pl}}(c) = 2v_{\text{pl}}(\gamma) = (0 \text{ or } 2) \bmod 4, \tag{S15}$$

with $v_{\text{pl}}(\gamma) = 0$ or $1 \in \pi_1(X, X_\Pi) = \mathbb{Z}_2$. Comparing with Eq. (S4), we know that if the $y$-mirror symmetry is broken, both the trivial and nontrivial phases in Eq. (S14) will reduce to the trivial phase. Therefore, the mirror symmetry is crucial to preventing the double-twist strip from unknotting.

## 2. Relation between mirror-symmetry protected $\mathbb{Z}_2$ topology and C lines

Now we discuss the relation of the twist of the polarization vectors along the semi-loops, and the number of the C lines at one side of the mirror plane enclosed by the semi-loops. First, we connect the two end points $\mathbf{r}_0, \mathbf{r}_1$ of the semi-loop $\gamma$ with an arbitrary curve $s_1 \subset \Pi$, and hence obtain a closed loop $\gamma_1 = \gamma \circ s_1$ (see Fig. S3(a)). Severing the loop $\gamma_1$ at the base point $\mathbf{r}_0$, we lift the major-axis bivectors and phase $(\mathbf{A}, e^{i\theta}) \sim (-\mathbf{A}, -e^{i\theta}) \in \frac{S^2 \times S^1}{\mathbb{Z}_2}$ to $(\mathbf{A}, e^{i\theta}) \in S^2 \times S^1$ with fixing a definite direction of $\mathbf{A}$ and a definite value of $\theta$ at each point along the severed path $\gamma'$ such that they are continuous along the path except for the breakpoint $\mathbf{r}_0$ (see Fig. S3(b)). Along $s_1$, $\mathbf{A}(\mathbf{r} \in s)$ remains fixed to $\hat{\mathbf{y}}$ or $-\hat{\mathbf{y}}$, i.e., we have $\frac{\mathbf{A}(k \in s)}{|\mathbf{A}(k \in s)|} \equiv \frac{\mathbf{A}_1}{|\mathbf{A}_1|}$ with $\mathbf{A}_1 = \mathbf{A}(\mathbf{r}_1)$. On the other hand, $\mathbf{A} e^{i\theta}$ is globally single-valued, $\mathbf{A}(\mathbf{r}_0) e^{i\theta(\mathbf{r}_0)} = \mathbf{A}_0 e^{i\theta_0} = \mathbf{A}_f e^{i\theta_f}$ where $(\mathbf{A}_0, e^{i\theta_0})$ and $(\mathbf{A}_f, e^{i\theta_f})$ denote the lift vectors and phases at the starting and end points of $\gamma'$ (see Fig. S3(b)). These two facts lead to the equality

$$\exp[i\pi I_{\text{ph}}(\gamma_1)] = \exp[i(\theta_f - \theta_0)] = \text{sign}(\mathbf{A}_0 \cdot \mathbf{A}_f) = \text{sign}(\mathbf{A}_0 \cdot \mathbf{A}_1) = \exp[i\pi v_{\text{pl}}(\gamma)] \quad (S16)$$

and equivalently,

$$v_{\text{pl}}(\gamma) = I_{\text{ph}}(\gamma_1) \mod 2. \quad (S17)$$

This result reveals that *(1) in the trivial twist phase of the major axis with $v_{pl}(\gamma) = 0$, the number of C lines on one side of $\Pi$ enclosed by the semi-loop $\gamma$ is even; (2) in the Möbius twist phase of the major axis with $v_{pl}(\gamma) = 1$, the number of C lines on one side of $\Pi$ enclosed by $\gamma$ is odd.* This relation is illustrated in Fig. S3(b) and (c) for nontrivial and trivial semi-loops, respectively.

We also note that the arc connecting the two terminal points can be arbitrarily selected. For two selections $s_1$ and $s_2$ (see Fig. S3(a)), the corresponding closed loops are $\gamma_1 = \gamma \circ s_1$ and $\gamma_2 = \gamma \circ s_2$. Their phase indices can be different, providing that the closed path $s_2^{-1} \circ s_1$ encircles some V points in the plane $\Pi$:

$$I_{\text{ph}}(\gamma_1) = I_{\text{ph}}(\gamma_2) + I_{\text{ph}}(s_2^{-1} \circ s_1). \quad (S18)$$

However, since $s_2^{-1} \circ s_1 \subset \Pi$ must carry an even phase index, the parity of the phase index along $\gamma \circ s$ for arbitrary $s$ are always invariant

$$I_{\text{ph}}(\gamma_1) = I_{\text{ph}}(\gamma_2) \mod 2, \quad (S19)$$

which is consistent with Eq. (S17) and confirms that the topological classification along the semi-loops is well-defined.

# Supplementary Note 4: Topological indices of the central C lines protected by generalized rotational symmetry

In this section, we examine the minimal stable charges of the C lines along $y$ axis protected by different discrete rotational symmetries $\bar{C}_n$. Consider a circle, $c$, which is centered on the $y$ axis and has the positive normal direction along $+\hat{y}$. In terms of the $n$-fold generalized rotational symmetry $\bar{C}_n$, the scalar field $\Psi = \mathbf{H} \cdot \mathbf{H}$ satisfies $\Psi\left(R\left(\frac{2\pi}{n}\right)\mathbf{r}\right) = e^{\frac{i4\pi}{n}}\Psi(\mathbf{r})$. Using this relation, we obtain the $n\mathbb{Z}$-quantized phase index along the circle $c$:

$$\begin{aligned}I_{\text{ph}}(c) &= \frac{1}{2\pi}\oint d\text{Arg}\Psi(R(\phi)\mathbf{r}_0) = \frac{n}{2\pi}\int_0^{\frac{2\pi}{n}}d\phi\, \partial_\phi \text{Arg}\Psi(R(\phi)\mathbf{r}_0)\\ &= \frac{n}{2\pi}\left[\text{Arg}\Psi\left(R\left(\frac{2\pi}{n}\right)\mathbf{r}_0\right) - \text{Arg}\Psi(\mathbf{r}_0)\right] = \frac{n}{2\pi}\left[\frac{4\pi}{n} + 2m\pi\right]\\ &= 2 + nm \in \{\cdots, 2-2n, 2-n, 2, 2+n, 2+2n, \cdots\},\end{aligned} \quad (S20)$$

where $\mathbf{r}_0$ is an arbitrary point on the circle. As we have analyzed in the main text, the minimal charge of $I_{\text{ph}}(c)$ determines the polarization and phase indices of the C line along the $y$ axis $I_{\text{pl}} = +\frac{I_{\text{ph}}^{\min}}{2}$. Therefore, for $n = 1, 2$, $I_{\text{pl}} = 0$, indicating the $y$ axis is not a topologically stable C line; for $n = 3$, $I_{\text{pl}} = -1$; for $n = 4$, $I_{\text{pl}} = \pm 1$, indicating the central C line is second order with either positive or negative charges; for $n \geq 5$, the central C lines is always of positive second order with $I_{\text{pl}} = +1$.

In what follows, we will demonstrate this general relation between the order of the central C line and the order of the discrete generalized rotational symmetry via the perturbation analysis near the $y$ axis. For convenience, we adopt the circular basis of transverse coordinates $r_\pm = z \pm ix = re^{\pm i\phi}$ which satisfies $R(\phi)r_\pm = e^{\pm i\phi}r_\pm$, and expand the field in the transverse plane (i.e., $xoz$-plane) of a fixed point $\mathbf{r}_0 = y_0\hat{y}$ on the $y$ axis:

$$\begin{aligned}\mathbf{H}(\mathbf{r}) &= \mathbf{H}(r_+, r_-, y_0) \\ &= \mathbf{h}^{(0)} + (\mathbf{h}_+^{(1)}r_+ + \mathbf{h}_-^{(1)}r_-) + (\mathbf{h}_{++}^{(2)}r_+^2 + \mathbf{h}_{--}^{(2)}r_-^2 + \mathbf{h}_{\pm}^{(2)}r_+r_-) + \mathcal{O}(r^3)\end{aligned} \quad (S21)$$

where $\mathbf{h}^{(0)} = \mathbf{H}(\mathbf{r}_0) = \mathbf{R}$. According to Eq. (6), the first order terms satisfy

$$e^{\mp\frac{i2\pi}{n}}R\left(\frac{2\pi}{n}\right)\mathbf{h}_\pm^{(1)} = e^{-\frac{i2\pi}{n}}\mathbf{h}_\pm^{(1)}$$

$$\Rightarrow \begin{cases} R\left(\frac{2\pi}{n}\right)\mathbf{h}_+^{(1)} = \mathbf{h}_+^{(1)} & \Rightarrow \quad \mathbf{h}_+^{(1)} = a_1\hat{y} \\ R\left(\frac{2\pi}{n}\right)\mathbf{h}_-^{(1)} = e^{-\frac{i4\pi}{n}}\mathbf{h}_-^{(1)} & \Rightarrow \quad \mathbf{h}_-^{(1)} = a_2\mathbf{L}\ (n=3)\ \text{or}\ \mathbf{0}\ (n \geq 4) \end{cases}. \quad (S22)$$

The second order terms satisfy

$$e^{-\frac{i4\pi}{n}}R\left(\frac{2\pi}{n}\right)\mathbf{h}_{++}^{(2)} = e^{-\frac{i2\pi}{n}}\mathbf{h}_{++}^{(2)} \Rightarrow R\left(\frac{2\pi}{n}\right)\mathbf{h}_{++}^{(2)} = e^{\frac{i2\pi}{n}}\mathbf{h}_{++}^{(2)} \Rightarrow \mathbf{h}_{++}^{(2)} = a_3\mathbf{L} \quad (S23)$$

$$e^{\frac{i4\pi}{n}}R\left(\frac{2\pi}{n}\right)\mathbf{h}_{--}^{(2)} = e^{-\frac{i2\pi}{n}}\mathbf{h}_{--}^{(2)} \Rightarrow R\left(\frac{2\pi}{n}\right)\mathbf{h}_{--}^{(2)} = e^{-\frac{i6\pi}{n}}\mathbf{h}_{--}^{(2)}$$

$$\Rightarrow \mathbf{h}_{--}^{(2)} = a_4\hat{\mathbf{y}}\ (n=3) \text{ or } a_4\mathbf{L}\ (n=4) \text{ or } \mathbf{0}\ (n \geq 5) \quad (S24)$$

$$R\left(\frac{2\pi}{n}\right)\mathbf{h}_{\pm}^{(2)} = e^{-\frac{i2\pi}{n}}\mathbf{h}_{\pm}^{(2)} \Rightarrow \mathbf{h}_{\pm}^{(2)} = a_5\mathbf{R}. \quad (S25)$$

Therefore, we obtain the perturbative magnetic field near the $y$ axis under the constrains of different $\bar{C}_n$ symmetry.

$\bar{C}_3$ symmetry case:

$$\begin{aligned}\mathbf{H}(r_+, r_-, z_0) &= \mathbf{R} + (a_1\hat{\mathbf{y}}r_+ + a_2\mathbf{L}r_-) + (a_3\mathbf{L}r_+^2 + a_4\hat{\mathbf{y}}r_-^2 + a_5\mathbf{R}r_+r_-) + \mathcal{O}(r^3)\\ &= \mathbf{R} + r(a_1\hat{\mathbf{y}}e^{i\phi} + a_2\mathbf{L}e^{-i\phi}) + r^2(a_3\mathbf{L}e^{i2\phi} + a_4\hat{\mathbf{z}}e^{-i2\phi} + a_5\mathbf{R}) + \mathcal{O}(r^3)\end{aligned} \quad (S26)$$

$$\Rightarrow \Psi(r,\phi) = \mathbf{H}(r,\phi)\cdot\mathbf{H}(r,\phi) = 2a_2 re^{-i\phi} + \mathcal{O}(r^2). \quad (S27)$$

Therefore, by $\bar{C}_3$ symmetry, the C-line along the $y$ axis takes stable charge $I_{\text{ph}} = -1$.

$\bar{C}_4$ symmetry case:

$$\begin{aligned}\mathbf{H}(r_+, r_-, z_0) &= \mathbf{R} + a_1\hat{\mathbf{y}}r_+ + (a_3\mathbf{L}r_+^2 + a_4\mathbf{L}r_-^2 + a_5\mathbf{R}r_+r_-) + \mathcal{O}(r^3)\\ &= \mathbf{R} + ra_1\hat{\mathbf{y}}e^{i\phi} + r^2(a_3\mathbf{L}e^{i2\phi} + a_4\mathbf{L}e^{-i2\phi} + a_5\mathbf{R}) + \mathcal{O}(r^3)\end{aligned} \quad (S28)$$

$$\Rightarrow \Psi(r,\phi) = \mathbf{H}(r,\phi)\cdot\mathbf{H}(r,\phi) = r^2[(a_1^2 + a_3)e^{2i\phi} + a_4 e^{-2i\phi}] + \mathcal{O}(r^3). \quad (S29)$$

Therefore, we conform that $\bar{C}_4$ symmetry requires the charge of the central C line takes either $I_{\text{ph}} = 2$ (if $|a_1^2 + a_3| > |a_4|$) or $I_{\text{ph}} = -2$ (if $|a_1^2 + a_3| < |a_4|$). When $|a_1^2 + a_3| = |a_4|$, we have $\Psi\left(r, \frac{1}{4}(\pi + \arg(\frac{a_1^2+a_3}{a_4})) + \frac{n\pi}{2}\right) = 0$ which indicates four first-order C lines grow out from the central C-line at the angles $\phi_n = \frac{1}{4}(\pi + \arg(\frac{a_1^2+a_3}{a_4})) + \frac{n\pi}{2}, (n = 0, 1, 2, 3)$.

$\bar{C}_n\ (n \geq 5)$ symmetry case:

$$\begin{aligned}\mathbf{H}(r_+, r_-, z_0) &= \mathbf{R} + a_1\hat{\mathbf{y}}r_+ + (a_3\mathbf{L}r_+^2 + a_5\mathbf{R}r_+r_-) + \mathcal{O}(r^3)\\ &= \mathbf{R} + ra_1\hat{\mathbf{y}}e^{i\phi} + r^2(a_3\mathbf{L}e^{i2\phi} + a_5\mathbf{R}) + \mathcal{O}(r^3)\end{aligned} \quad (S30)$$

$$\Rightarrow \Psi(r,\phi) = \mathbf{H}(r,\phi)\cdot\mathbf{H}(r,\phi) = r^2(a_1^2 + a_3)e^{2i\phi} + \mathcal{O}(r^3). \quad (S31)$$

Therefore, as long as the degree of the $\bar{C}_n$ symmetry $n \geq 5$, the central C line always takes the charge $I_{\text{ph}} = +2$.

The minimal charge of the central C line protected by $\bar{C}_n$ symmetry also determines the stable polarization patterns in the vicinity of the central axis. In Fig. S4, using Eqs. (S26), (S28) and (S30), we depict four representative structures of the major axis streamlines in the transverse $xoz$-plane near the central C line in the presence of different $\bar{C}_n$ symmetries. It can be seen that $\bar{C}_n$ symmetry requires the patterns of the major axis are $C_n$ symmetric. For example, the three-pointed star pattern with $I_{pl} = -1/2$ is the only possible $C_3$ symmetric configuration with $|I_{pl}| = 1/2$, and all other $C_3$ symmetric singular patterns have higher $|I_{pl}|$ and cannot stably exit without more strict symmetry constraints. Similarly, all patterns of $|I_{pl}| = 1/2$ are incompatible with any $\bar{C}_n$ ($n \geq 4$) symmetry. The spiral vortex ($I_{pl} = 1$) and the saddle ($I_{pl} = -1$) are the only two patterns with $|I_{pl}| = 1$ admitted by $\bar{C}_4$ symmetry. And for $\bar{C}_{n \geq 5}$ (including the cylindrical case), the spiral vortex is the pattern with minimal polarization index and compatible with the rotational symmetry. When the chirality of incident wave is changed to be left-handed, the generalized rotation symmetries change to $R(\frac{2\pi}{n})\mathbf{H}(R(-\frac{2\pi}{n})\mathbf{r}) = e^{+\frac{i2\pi}{n}}\mathbf{H}(\mathbf{r})$, which also requires the patterns of major axis near the central axis are $C_n$ symmetric. Consequently, the minimal polarization index of the central C line is irrelevant to the chirality of the circularly polarized incident field but only depends on the rotation symmetry of the scatters.

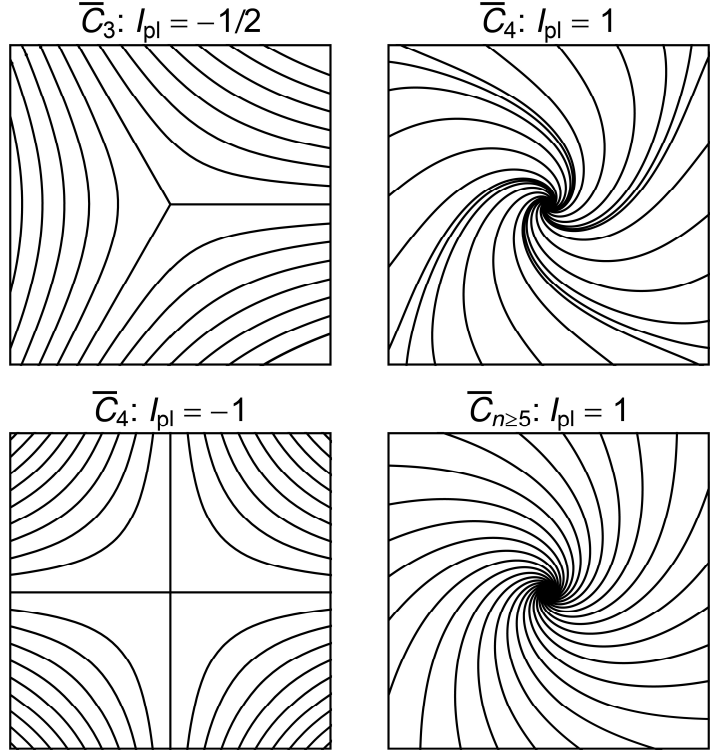

**Fig. S4**. Streamlines of major axis in the transverse plane near the stable central C points protected by $\bar{C}_n$ symmetries.

# Reference

[1] J. D. Jackson, *Classical Electrodynamics*, 3 edition (Wiley, New York, 1998).

[2] J. F. Nye and J. V. Hajnal, *The Wave Structure of Monochromatic Electromagnetic Radiation*, Proc. R. Soc. Lond. A **409**, 21 (1987).

[3] K. Y. Bliokh, M. A. Alonso, and M. R. Dennis, *Geometric Phases in 2D and 3D Polarized Fields: Geometrical, Dynamical, and Topological Aspects*, Rep. Prog. Phys. **82**, 122401 (2019).

[4] M. V. Berry, *Index Formulae for Singular Lines of Polarization*, J. Opt. A: Pure Appl. Opt. **6**, 675 (2004).

[5] F. Flossmann, U. T. Schwarz, M. Maier, and M. R. Dennis, *Polarization Singularities from Unfolding an Optical Vortex through a Birefringent Crystal*, Phys. Rev. Lett. **95**, 253901 (2005).

[6] A. Hatcher, *Algebraic Topology*, 1st edition (Cambridge University Press, New York, 2001).

[7] T. Bauer, P. Banzer, E. Karimi, S. Orlov, A. Rubano, L. Marrucci, E. Santamato, R. W. Boyd, and G. Leuchs, *Observation of Optical Polarization Möbius Strips*, Science **347**, 964 (2015).

[8] I. Freund, *Optical Möbius Strips in Three-Dimensional Ellipse Fields: I. Lines of Circular Polarization*, Opt. Commun. **283**, 1 (2010).

[9] I. Freund, *Multitwist Optical Möbius Strips*, Opt. Lett. **35**, 148 (2010).

[10] E. J. Galvez, I. Dutta, K. Beach, J. J. Zeosky, J. A. Jones, and B. Khajavi, *Multitwist Möbius Strips and Twisted Ribbons in the Polarization of Paraxial Light Beams*, Sci. Rep. **7**, 13653 (2017).

[11] C. C. Wojcik, X.-Q. Sun, T. Bzdušek, and S. Fan, *Homotopy Characterization of Non-Hermitian Hamiltonians*, Phys. Rev. B **101**, 205417 (2020).

[12] X.-Q. Sun, S.-C. Zhang, and T. Bzdušek, *Conversion Rules for Weyl Points and Nodal Lines in Topological Media*, Phys. Rev. Lett. **121**, 106402 (2018).